\documentclass[reprint,aps,prb,longbibliography]{revtex4-1}

\usepackage{graphicx}
\usepackage{amsmath}
\usepackage{bm}
\usepackage{amssymb}
\usepackage{amsfonts}
\usepackage{color}
\usepackage{graphicx}
\usepackage{epstopdf}
\usepackage{braket}

\begin{document}

\title{The effect of microscopic scattering on the nonlinear transmission of terahertz fields through monolayer graphene}

\author{L. G. Helt}

\affiliation{Department of Physics, Engineering Physics and Astronomy, Queen's University, Kingston, Ontario K7L 3N6, Canada}

\author{M. M. Dignam}

\affiliation{Department of Physics, Engineering Physics and Astronomy, Queen's University, Kingston, Ontario K7L 3N6, Canada}

\begin{abstract}
We consider the nonlinear terahertz response of n-doped monolayer graphene at room temperature using a microscopic theory of carrier dynamics. Our tight-binding model treats the carrier-field interaction in the length gauge, includes phonon as well as short-range neutral-impurity scattering, and fully accounts for the intrinsic nonlinear response of graphene near the Dirac point. Treating each interaction microscopically allows us to separate contributions from current clipping, phonon creation, and elastic impurity scattering. Although neutral impurity scattering and phonon scattering are both highly energy-dependent, we find that they impact conduction-band electron dynamics very differently, and that together they can help explain experimental results concerning field-dependent terahertz transmission through graphene.
\end{abstract}

\maketitle

\section{Introduction}
Graphene, a monolayer of carbon atoms arranged in a hexagonal lattice, has attracted much interest since it was first isolated using the Scotch Tape method~\cite{Novoselov:2004}. Indeed, its unique electronic bandstructure has led to proposed applications including photovoltaics, display panels, and sensors, and research is slowly beginning to move from the laboratory to industry~\cite{Randvir:2014,Zhu:2018}. Nurturing this move, terahertz (THz) spectroscopy has emerged as a tool well-suited to characterize graphene carrier dynamics~\cite{Maeng:2012,Ren:2012,Paul:2013,Hafez:2014,Tomadin:2018}. 

THz spectroscopy is fast and non-destructive, indeed, non-contact, and able to monitor fabrication in situ. It can reveal carrier dynamics, and is ideal for exploring the bandstructure of graphene near the Dirac point because THz frequencies are low enough to drive intraband dynamics yet also high enough to cause interband transitions in undoped graphene~\cite{Al-Naib:2014,Al-Naib:2015}. At large field amplitudes THz spectroscopy can even be used to study nonlinear dynamics, as the intraband current can become saturated or ``clipped'' due to the nonlinear relationship between carrier crystal momentum and carrier velocity~\cite{Hafez:2014}. Particularly worthy of investigation at these high THz field amplitudes, where a simple Drude model breaks down, are the effects of phonon and impurity scattering on the nonlinear dynamics of carriers in graphene and their impact on field-dependent THz transmission.

\begin{figure}[htbp]
\centering
\includegraphics[width=0.9\linewidth]{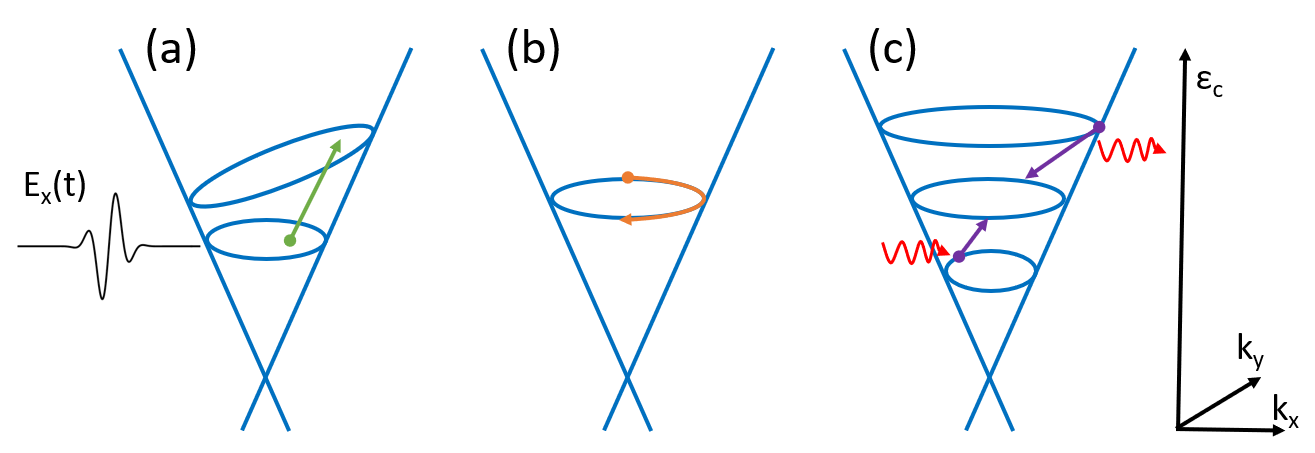}
\caption{Sketch of the intraband processes we consider near the Dirac point and their effects. (a) Carrier-field interactions, with negative $E_{x}$ amplitudes driving carriers to larger $k_{x}$. (b) Carrier-neutral impurity interactions, which are elastic in carrier energy but redistribute carrier momentum. (c) Carrier-optical phonon interactions, which can reduce (increase) carrier energy and redistribute momentum upon creation (annihilation) of a phonon.}
\label{fig:Processes}
\end{figure}

In this work we theoretically investigate the interplay between intrinsic current saturation, phonon scattering, and neutral impurity scattering in n-doped graphene (see Fig.~\ref{fig:Processes}) with increasing THz field strength. We employ a microscopic theory of carriers and their interactions under the tight-binding approximation and in the length-gauge. While others have explored the nonlinear response of graphene at THz frequencies using empirical models of scattering~\cite{Al-Naib:2014,Al-Naib:2015} and at \textit{optical} frequencies using microscopic models~\cite{Winnerl:2011,Malic:2011,Winnerl:2017, Malic:2017}, to our knowledge this is the first work to examine the nonlinear THz response of graphene that includes a microscopic model of carrier scattering.

In Section~\ref{sec:Model} we develop our model. We consider n-doped graphene here for simplicity, allowing us to focus on intraband dynamics in the conduction band. We take the Fermi level high enough that THz pulses do not excite interband transitions and optical phonon scattering rates dominate over acoustic phonon scattering rates~\cite{Fang:2011}. We include short-range neutral impurity scattering, which can often dominate over charged-impurity scattering for graphene on standard substrates~\cite{Razavipour:2015}. For simplicity, in this work we neglect carrier-carrier scattering. In Section~\ref{sec:Results} we use our model to simulate carrier dynamics, currents, transmitted fields, and conductivities as functions of incident field strength. To help unravel the role that each scattering channel plays in intraband dynamics, we first consider artificial scenarios, in which no scattering or only a single scattering mechanism is present, before performing a full situation where both phonon and impurity scattering are present. The carrier dynamics and nonlinear conductivities are compared against an earlier semi-empirical model of scattering~\cite{Al-Naib:2014,Al-Naib:2015}. We then use our full model in Section~\ref{sec:Application} to simulate the field-dependent transmission of graphene and compare our results to experimental THz time-domain spectroscopy (TDS) measurements demonstrating nonlinear THz transmission through graphene~\cite{Hafez:2014}. Finally, we conclude in Section~\ref{sec:Conclusion}.

\section{Theory}
\label{sec:Model}
We calculate the conduction and valence bands of graphene in the nearest-neighbor tight-binding approximation~\cite{Al-Naib:2014,Al-Naib:2015}. The field-carrier interaction is treated in the length-gauge, thus avoiding divergences at low frequencies~\cite{Al-Naib:2014,Aversa:1995} (such as the THz range). While in our group's previous work, semi-empirical models were employed to treat carrier scattering, in this paper we treat scattering at the microscopic level, including interactions of carriers with neutral impurities and optical phonons. We do not approximate dynamic equations to lowest order in the applied field~\cite{Ando:2006,Hwang:2008}, but allow for pump fields to drive carriers far from equilibrium. 

In addition to the carrier and carrier-field interaction Hamiltonians of Refs.~\onlinecite{Al-Naib:2014,Aversa:1995} we add the carrier-phonon Hamiltonian
\begin{align}
H_{\text{c-ph}}=&\sum_{n,m,\mathbf{k},\mathbf{q}}\left[g_{\mathbf{q}j}^{\mathbf{k}nm}a_{n}^{\dagger}\left(\mathbf{k}+\mathbf{q}\right)a_{m}\left(\mathbf{k}\right)b_{j}\left(\mathbf{q}\right)\right.\nonumber \\ 
&+\left.\left(g_{\mathbf{q}j}^{\mathbf{k}nm}\right)^{*}a_{m}^{\dagger}\left(\mathbf{k}\right)a_{n}\left(\mathbf{k}+\mathbf{q}\right)b_{j}^{\dagger}\left(\mathbf{q}\right)\right],
\end{align}
the phonon Hamiltonian
\begin{equation}
H_{\text{ph}}=\sum_{\mathbf{q}}\hbar\omega_{j}\left(\mathbf{q}\right)b_{j}^{\dagger}\left(\mathbf{q}\right)b_{j}\left(\mathbf{q}\right),
\end{equation}
and the carrier-neutral impurity Hamiltonian
\begin{equation}
H_{\text{c-i}}=\sum_{n,\mathbf{k},\mathbf{q}}h_{\mathbf{q}}^{\mathbf{k}n}a_{n}^{\dagger}\left(\mathbf{q}\right)a_{n}\left(\mathbf{k}\right).
\end{equation}
Here the $a_{n}\left(\mathbf{k}\right)$ are fermionic carrier operators labeled by a band index $n$ and Bloch wave vector $\mathbf{k}$, the $b_{j}\left(\mathbf{q}\right)$ are bosonic phonon operators labeled by phonon branch index $j$ (e.g. longitudinal optical) and wave vector $\mathbf{q}$, $\hbar\omega_{j}\left(\mathbf{q}\right)$ is the phonon energy, $g_{\mathbf{q}j}^{\mathbf{k}nm}$ a carrier-phonon coupling element (defined below), and $h_{\mathbf{q}}^{\mathbf{k}n}$ a carrier-neutral impurity coupling element (also defined below).

In this work, we consider only doped graphene with a chemical potential $\mu_{c}$ that is high $\left(>100\text{ meV}\right)$ in the conduction band. Because we will be considering exciting pulses with a central frequency of only 1 THz (photon energy of $\sim 4 \text{ meV}$), this allows us to ignore interband transitions and valence band electrons, including only electrons in the conduction band. Calculating scattering dynamics up to the second-order Born-Markov approximation~\cite{Kuhn:1998,Rossi:2002}, the resulting dynamic equation for the conduction band electron population is
\begin{align}
\frac{\text{d}\rho_{cc}\left(\mathbf{k}\right)}{\text{d}t}=&-\frac{e\mathbf{E}^{t}\left(t\right)}{\hbar}\cdot\nabla_{\mathbf{k}}\rho_{cc}\left(\mathbf{k}\right)\nonumber \\
&+\frac{i}{\hbar}\left\langle\left[H_{\text{c-ph}}+H_{\text{ph}}+H_{\text{c-i}},a_{c}^{\dagger}\left(\mathbf{k}\right)a_{c}\left(\mathbf{k}\right)\right]\right\rangle\nonumber \\
=&-\frac{e\mathbf{E}^{t}\left(t\right)}{\hbar}\cdot\nabla_{\mathbf{k}}\rho_{cc}\left(\mathbf{k}\right)-\Gamma_{c}^{\text{out}}\left(\mathbf{k}\right)\rho_{cc}\left(\mathbf{k}\right)\nonumber \\
&+\Gamma_{c}^{\text{in}}\left(\mathbf{k}\right)\left[1-\rho_{cc}\left(\mathbf{k}\right)\right],
\label{eq:SBE}
\end{align}
where
\begin{align}
\Gamma_{c}^{\text{out}}\left(\mathbf{k}\right)=&\frac{2\pi}{\hbar}\sum_{\mathbf{q}j}\left\{\left|g_{\mathbf{q}j}^{\mathbf{k}cc}\right|^{2}\left[1-\rho_{cc}\left(\mathbf{k}+\mathbf{q}\right)\right]n_{j}\left(\mathbf{q}\right)\right.\nonumber\\
&\quad\quad\quad\quad\times\delta\left[\varepsilon_{c}\left(\mathbf{k}+\mathbf{q}\right)-\varepsilon_{c}\left(\mathbf{k}\right)-\hbar\omega_{j}\left(\mathbf{q}\right)\right]\nonumber\\
&\quad\quad\quad+\left|g_{\mathbf{q}j}^{\mathbf{k}-\mathbf{q}cc}\right|^{2}\left[1-\rho_{cc}\left(\mathbf{k}-\mathbf{q}\right)\right]\left(n_{j}\left(\mathbf{q}\right)+1\right)\nonumber\\
&\quad\quad\quad\quad\times\delta\left[\varepsilon_{c}\left(\mathbf{k}-\mathbf{q}\right)-\varepsilon_{c}\left(\mathbf{k}\right)+\hbar\omega_{j}\left(\mathbf{q}\right)\right]\nonumber\\
&\quad\quad\quad+\left.\left|h_{\mathbf{q}}^{\mathbf{k}c}\right|^{2}\left[1-\rho_{cc}\left(\mathbf{q}\right)\right]\delta\left[\varepsilon_{c}\left(\mathbf{q}\right)-\varepsilon_{c}\left(\mathbf{k}\right)\right]\right\},
\end{align}
is the scattering-out rate and
\begin{align}
\Gamma_{c}^{\text{in}}\left(\mathbf{k}\right)=&\frac{2\pi}{\hbar}\sum_{\mathbf{q}j}\left\{\left|g_{\mathbf{q}j}^{\mathbf{k}cc}\right|^{2}\rho_{cc}\left(\mathbf{k}+\mathbf{q}\right)\left(n_{j}\left(\mathbf{q}\right)+1\right)\right.\nonumber\\
&\quad\quad\quad\quad\times\delta\left[\varepsilon_{c}\left(\mathbf{k}+\mathbf{q}\right)-\varepsilon_{c}\left(\mathbf{k}\right)-\hbar\omega_{j}\left(\mathbf{q}\right)\right]\nonumber\\
&\quad\quad\quad+\left|g_{\mathbf{q}j}^{\mathbf{k}-\mathbf{q}cc}\right|^{2}\rho_{cc}\left(\mathbf{k}-\mathbf{q}\right)n_{j}\left(\mathbf{q}\right)\nonumber\\
&\quad\quad\quad\quad\times\delta\left[\varepsilon_{c}\left(\mathbf{k}-\mathbf{q}\right)-\varepsilon_{c}\left(\mathbf{k}\right)+\hbar\omega_{j}\left(\mathbf{q}\right)\right]\nonumber\\
&\quad\quad\quad+\left.\left|h_{\mathbf{q}}^{\mathbf{k}c}\right|^{2}\rho_{cc}\left(\mathbf{q}\right)\delta\left[\varepsilon_{c}\left(\mathbf{q}\right)-\varepsilon_{c}\left(\mathbf{k}\right)\right]\right\},
\end{align}
the scattering-in rate. Here $\mathbf{E}^{t}\left(t\right)$ is the (transmitted) electric field at the graphene, $\rho_{cc}\left(\mathbf{k}\right)=\left\langle a_{c}^{\dagger}\left(\mathbf{k}\right)a_{c}\left(\mathbf{k}\right)\right\rangle$ the conduction band density matrix, $n_{j}\left(\mathbf{q}\right)=\left\langle b_{j}^{\dagger}\left(\mathbf{q}\right)b_{j}\left(\mathbf{q}\right)\right\rangle$ the phonon population, and $\varepsilon_{c}\left(\mathbf{k}\right)$ the electron energy.

As we work near the Dirac point, we take the electron dispersion relation to be linear, $\varepsilon_{c}\left(\mathbf{k}\right)=\hbar v_{\text{F}}\left\vert\mathbf{k}\right\vert$, where $v_{\text{F}}=9.81\times 10^5\text{ m/s}$ is the Fermi velocity. We include longitudinal optical and transverse optical phonon modes near the $\Gamma$ point ($j=\Gamma\text{-LO},\Gamma\text{-TO}$) as well as the phonon modes near the $K$ point ($j=K$), and take the phonon energies to be dispersionless ($\hbar\omega_{\Gamma-\text{LO}}\left(\mathbf{q}\right)\approx\hbar\omega_{\Gamma-\text{TO}}\left(\mathbf{q}\right)\approx\hbar\omega_{\Gamma}=196\text{ meV}$, $\hbar\omega_{K}\left(\mathbf{q}\right)\approx\hbar\omega_{K}=160\text{ meV}$)~\cite{Fang:2011}. As a first approximation, valid for the THz pulse durations and field strengths we consider, we do not include any phonon dynamics, treating the phonons of interest as thermal baths at constant temperature. This allows us to write the equilibrium phonon population in a given mode as $n_{\Gamma}=1/\left(\exp\left[\hbar\omega_{\Gamma}/\left(k_{\text{B}}T\right)\right]-1\right)$, and $n_{K}=1/\left(\exp\left[\hbar\omega_{K}/\left(k_{\text{B}}T\right)\right]-1\right)$, where $T$ is the (initial) carrier temperature and $k_{\text{B}}$ is the Boltzmann constant. 

The required carrier-phonon coupling elements are~\cite{Malic:2011}
\begin{align}
\left|g_{\mathbf{q}\Gamma\text{-LO}}^{\mathbf{k}cc}\right|^{2}=&\frac{1}{N}g_{\Gamma}^{2}\left[1-\cos\left(\theta_{\mathbf{q},\mathbf{k}}+\theta_{\mathbf{q},\mathbf{k}+\mathbf{q}}\right)\right]\nonumber \\
\left|g_{\mathbf{q}\Gamma\text{-TO}}^{\mathbf{k}cc}\right|^{2}=&\frac{1}{N}g_{\Gamma}^{2}\left[1+\cos\left(\theta_{\mathbf{q},\mathbf{k}}+\theta_{\mathbf{q},\mathbf{k}+\mathbf{q}}\right)\right]\nonumber \\
\left|g_{\mathbf{q}K}^{\mathbf{k}cc}\right|^{2}=&\frac{1}{N}g_{K}^{2}\left[1-\cos\left(\theta_{\mathbf{k},\mathbf{k}+\mathbf{q}}\right)\right],
\end{align}
where $g_{\Gamma}^{2}=\text{ 0.0405 eV}^{2}$, $g_{K}^{2}=\text{ 0.0994 eV}^{2}$, $\theta_{\mathbf{k},\mathbf{q}}$ is the angle between $\mathbf{k}$ and $\mathbf{q}$, and $N$ the number of unit cells. The carrier-neutral impurity coupling elements are~\cite{Hwang:2008}
\begin{equation}
\left|h_{\mathbf{q}}^{\mathbf{k}c}\right|^{2}=\frac{n_{i}v_{0}^{2}}{A}\left[1+\cos\left(\theta_{\mathbf{k},\mathbf{q}}\right)\right],
\end{equation}
where $n_{i}$ is the neutral impurity density,  $v_{0}$ a constant interaction strength as appropriate for short-range point defect scatterers, and $A$ the area of the graphene sheet. All of this allows us to rewrite the scattering-in and -out rates as
\begin{align}
\Gamma_{c}^{\text{out}}\left(\mathbf{k}\right)=&\frac{2\pi}{\hbar N}\sum_{\mathbf{q}}\left[1-\rho_{cc}\left(\mathbf{q}\right)\right]\nonumber\\
&\times\left\{\vphantom{\frac{N n_{i}v_{0}^{2}}{A}}2g_{\Gamma}^{2}n_{\Gamma}\delta\left[\varepsilon_{c}\left(\mathbf{q}\right)-\varepsilon_{c}\left(\mathbf{k}\right)-\hbar\omega_{\Gamma}\right]\right.\nonumber\\
&\quad+2g_{\Gamma}^{2}\left(n_{\Gamma}+1\right)\delta\left[\varepsilon_{c}\left(\mathbf{q}\right)-\varepsilon_{c}\left(\mathbf{k}\right)+\hbar\omega_{\Gamma}\right]\nonumber\\
&\quad+g_{K}^{2}\left[1-\cos\left(\theta_{\mathbf{k},\mathbf{q}}\right)\right]n_{K}\nonumber\\
&\quad\quad\times\delta\left[\varepsilon_{c}\left(\mathbf{q}\right)-\varepsilon_{c}\left(\mathbf{k}\right)-\hbar\omega_{K}\right]\nonumber\\
&\quad+g_{K}^{2}\left[1-\cos\left(\theta_{\mathbf{k},\mathbf{q}}\right)\right]\left(n_{K}+1\right)\nonumber\\
&\quad\quad\times\delta\left[\varepsilon_{c}\left(\mathbf{q}\right)-\varepsilon_{c}\left(\mathbf{k}\right)+\hbar\omega_{K}\right]\nonumber\\
&\quad+\left.\frac{N n_{i}v_{0}^{2}}{A}\left[1+\cos\left(\theta_{\mathbf{k},\mathbf{q}}\right)\right]\delta\left[\varepsilon_{c}\left(\mathbf{q}\right)-\varepsilon_{c}\left(\mathbf{k}\right)\right]\right\},
\end{align}
and
\begin{align}
\Gamma_{c}^{\text{in}}\left(\mathbf{k}\right)=&\frac{2\pi}{\hbar N}\sum_{\mathbf{q}}\rho_{cc}\left(\mathbf{q}\right)\nonumber\\
&\times\left\{\vphantom{\frac{N n_{i}v_{0}^{2}}{A}}2g_{\Gamma}^{2}\left(n_{\Gamma}+1\right)\delta\left[\varepsilon_{c}\left(\mathbf{q}\right)-\varepsilon_{c}\left(\mathbf{k}\right)-\hbar\omega_{\Gamma}\right]\right.\nonumber\\
&\quad+2g_{\Gamma}^{2}n_{\Gamma}\delta\left[\varepsilon_{c}\left(\mathbf{q}\right)-\varepsilon_{c}\left(\mathbf{k}\right)+\hbar\omega_{\Gamma}\right]\nonumber\\
&\quad+g_{K}^{2}\left[1-\cos\left(\theta_{\mathbf{k},\mathbf{q}}\right)\right]\left(n_{K}+1\right)\nonumber\\
&\quad\quad\times\delta\left[\varepsilon_{c}\left(\mathbf{q}\right)-\varepsilon_{c}\left(\mathbf{k}\right)-\hbar\omega_{K}\right]\nonumber\\
&\quad+g_{K}^{2}\left[1-\cos\left(\theta_{\mathbf{k},\mathbf{q}}\right)\right]n_{K}\nonumber\\
&\quad\quad\times\delta\left[\varepsilon_{c}\left(\mathbf{q}\right)-\varepsilon_{c}\left(\mathbf{k}\right)+\hbar\omega_{K}\right]\nonumber\\
&\quad+\left.\frac{N n_{i}v_{0}^{2}}{A}\left[1+\cos\left(\theta_{\mathbf{k},\mathbf{q}}\right)\right]\delta\left[\varepsilon_{c}\left(\mathbf{q}\right)-\varepsilon_{c}\left(\mathbf{k}\right)\right]\right\}.
\end{align}
Note that one can obtain $\Gamma_{c}^{\text{out}}\left(\mathbf{k}\right)$ from $\Gamma_{c}^{\text{in}}\left(\mathbf{k}\right)$ by exchanging $\rho_{cc}\left(\mathbf{q}\right)\leftrightarrow 1-\rho_{cc}\left(\mathbf{q}\right)$ and $n_{j}\leftrightarrow n_{j}+1$. In light of this, for brevity, we only present expressions for $\Gamma_{c}^{\text{in}}\left(\mathbf{k}\right)$ below. We evaluate the two-dimensional sum over $\mathbf{q}$ as the integrals $\sum_{\mathbf{q}}\rightarrow\frac{A}{\left(2\pi\right)^{2}}\int_{0}^{\infty}\text{d}q\,q\int_{0}^{2\pi}\text{d}\theta_{q}$, and use the Dirac delta functions to perform integration over $q=\left\vert\mathbf{q}\right\vert$ to obtain
\begin{align}
&\Gamma_{c}^{\text{in}}\left(\mathbf{k}\right)\nonumber\\
&=\frac{A}{2\pi\hbar^{2}v_{\text{F}}N}\int_{0}^{2\pi}\text{d}\theta_{q}\nonumber\\
&\times\left\{\vphantom{\frac{N n_{i}v_{0}^{2}}{A}}2g_{\Gamma}^{2}\left(n_{\Gamma}+1\right)k_{\Gamma}^{+}\rho_{cc}\left[\mathbf{k}_{\Gamma}^{+}\left(\theta_{q}\right)\right]\right.\nonumber\\
&\quad+2g_{\Gamma}^{2}n_{\Gamma}k_{\Gamma}^{-}\rho_{cc}\left[\mathbf{k}_{\Gamma}^{-}\left(\theta_{q}\right)\right]\nonumber\\
&\quad+g_{K}^{2}\left[1-\cos\left(\theta_{\mathbf{k},\mathbf{k}_{K}^{+}\left(\theta_{q}\right)}\right)\right]\left(n_{K}+1\right)k_{K}^{+}\rho_{cc}\left[\mathbf{k}_{K}^{+}\left(\theta_{q}\right)\right]\nonumber\\
&\quad+g_{K}^{2}\left[1-\cos\left(\theta_{\mathbf{k},\mathbf{k}_{K}^{-}\left(\theta_{q}\right)}\right)\right]n_{K}k_{K}^{-}\rho_{cc}\left[\mathbf{k}_{K}^{-}\left(\theta_{q}\right)\right]\nonumber\\
&\quad+\left.\frac{N n_{i}v_{0}^{2}}{A}\left[1+\cos\left(\theta_{\mathbf{k},\mathbf{k}\left(\theta_{q}\right)}\right)\right]\left\vert\mathbf{k}\right\vert\rho_{cc}\left[\mathbf{k}\left(\theta_{q}\right)\right]\right\},
\label{eq:GammaIn}
\end{align}
where we have introduced
\begin{equation}
k_{\Gamma\left(K\right)}^{\pm}=\left|\mathbf{k}\right|\pm\frac{\omega_{\Gamma\left(K\right)}}{v_{\text{F}}},
\end{equation}
\begin{equation}
\mathbf{k}_{\Gamma\left(K\right)}^{\pm}\left(\theta\right)=k_{\Gamma\left(K\right)}^{\pm}\left(\cos\left(\theta\right)\hat{\mathbf{x}}+\sin\left(\theta\right)\hat{\mathbf{y}}\right),
\end{equation}
and
\begin{equation}
\mathbf{k}\left(\theta\right)=\left|\mathbf{k}\right|\left(\cos\left(\theta\right)\hat{\mathbf{x}}+\sin\left(\theta\right)\hat{\mathbf{y}}\right).
\end{equation}

We solve carrier dynamic equations by placing Eq.~\eqref{eq:SBE} on a triangular 451$\times$451-point grid~\cite{Grid} in $\mathbf{k}$ using a finite difference approximation, and solving the resulting differential equation numerically via a Runge-Kutta routine (see Ref.~\onlinecite{Al-Naib:2014}). For each ring of constant final momentum amplitude $\left\vert\mathbf{q}\right\vert$ needed to calculate the scattering-in and -out rates, where $\left\vert\mathbf{q}\right\vert$ is $k_{\Gamma}^{\pm}$, $k_{K}^{\pm}$, or $\left\vert\mathbf{k}\right\vert$, we discritize the integral over the angle $\theta_{q}$ as $\int_{0}^{2\pi}f\left(\theta_{q}\right)\text{d}\theta_{q}\rightarrow\Delta\theta_{q}\sum_{j=1}^{M_{q}}f\left(j\Delta\theta_{q}\right)$. Here $M_{q}=\text{nint}\left(2\pi\left\vert\mathbf{q}\right\vert/\Delta k\right)$, $\Delta\theta_{q}=2\pi/M_{q}$, $\Delta k$ is the spacing between gridpoints on the $\mathbf{k}$-space grid, and $\text{nint}$ denotes the nearest integer. In this manner, we keep the density of points on the ring approximately equal to the density of $\mathbf{k}$-space gridpoints. Finally, $A/N=3\sqrt{3}a_{0}^{2}/2$ is the unit cell area, where $a_{0}=1.42$ \AA~is the distance between nearest neighbor atoms~\cite{McGouran:2016}. To ensure that we do not lose any carriers in our simulation due to them being scattered or driven by the field outside of the simulation grid, we set the grid edge (i.e. the largest possible $\left\vert\mathbf{q}\right\vert$) as $1.5\times$ the maximum displacement of the edge of the electron disc when driven by the strongest incident field $\mathbf{E}^{i}\left(t\right)$ of interest when no scattering channels are present. That is, taking $\mathbf{E}^{i}\left(t\right)=E^{i}\left(t\right)\hat{\mathbf{x}}$, we set $\text{max}\left[\left|\mathbf{q}\right|\right]=1.5\times \left[-e\,\text{max}\left[\int_{0}^{t}\text{d}\tau\, E^{i}\left(\tau\right)\right]/\hbar+\mu_{c}/\left(v_{\text{F}}\hbar\right)\right]$. For this estimate, we use $E^{i}\left(t\right)$ rather than the $E^{t}\left(t\right)$ of Eq.~\eqref{eq:SBE} because it is available before performing simulations, and also because it will not lead to an underestimation of $\text{max}\left[\left|\mathbf{q}\right|\right]$, as $E^{i}\left(t\right)>E^{t}\left(t\right)$. Note that it is unlikely that the points specified for the density matrix along the ring $\rho_{cc}\left[\mathbf{q}\left(\theta_{q}\right)\right]$ will match up with those found in the carrier $\mathbf{k}$-grid used in Eq.~\eqref{eq:SBE}. Thus, we employ a simple bilinear interpolation scheme, first solving for the closest $\mathbf{k}$-space gridpoints, and then taking an appropriate weighted average. 

\section{Results and Discussion}
\label{sec:Results}
\begin{figure}[htbp]
\centering
\includegraphics[width=0.9\linewidth]{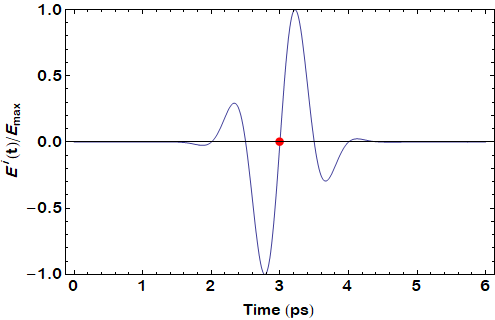}
\caption{Normalized incident field amplitude as a function of time. The time at which snapshots of $\rho_{cc}\left(\mathbf{k}\right)$ are taken is marked with a red circle.}
\label{fig:Ei}
\end{figure}

We take the incident field to be an approximately single-cycle pulse of the form
\begin{equation}
\mathbf{E}^{i}\left(t\right)=\frac{E_{\text{max}}}{N_{E}}e^{-\frac{4\ln\left(2\right)\left(t-t_{0}\right)^{2}}{\Delta_{t}^{2}}}\sin\left[2\pi f\left(t-t_{0}\right)\right]\hat{\mathbf{x}},
\label{eq:E}
\end{equation}
and the initial carrier population to be the thermal state
\begin{equation}
\rho_{cc}^{\text{init}}\left(\mathbf{k}\right)=\frac{1}{\exp\left[\left(\hbar v_{\text{F}}\left\vert\mathbf{k}\right\vert-\mu_{c}\right/\left(k_{\text{B}}T\right)\right]+1}.
\label{eq:rhot}
\end{equation}
Here $E_{\text{max}}$ is the field strength, $N_{E}$ a normalization constant defined below, $\Delta_{t}$ the pulse duration, $t_{0}$ a temporal offset, $f$ the pulse carrier wave frequency, and $T$ the initial temperature. In our calculations we consider $f=1.0\text{ THz}$, $\Delta_{t}=1.0\text{ ps}$, and $t_{0}=3.0\text{ ps}$, and we set $N_{E}=0.858937$ so that the peak of the field amplitude reaches $E_{\text{max}}$ at approximately $t=3.22\text{ ps}$ (see Fig.~\ref{fig:Ei}). We run simulations from $t=0\text{ ps}$ to $t=6.0\text{ ps}$ with $T=300\text{ K}$ and $\mu_{c}=354\text{ meV}$, giving an electron density of approximately $n_{c}=9.73\times 10^{12}\text{ cm}^{-2}$.

Throughout each run we calculate the intraband (here, the total) current as~\cite{McGouran:2016}
\begin{equation}
\mathbf{J}=e v_{\text{F}}\sum_{\mathbf{k}}\rho_{cc}\left(\mathbf{k}\right)\hat{\mathbf{k}},
\end{equation}
and self-consistently solve for the transmitted field
\begin{equation}
\mathbf{E}^{t}\left(t\right)=\frac{2\mathbf{E}^{i}\left(t\right)-Z_{0}\mathbf{J}\left[\mathbf{E}^{t}\left(t\right)\right]}{1+n},
\end{equation}
where $Z_{0}$ is the impedance of free space, $\mathbf{J}\left[\mathbf{E}^{t}\left(t\right)\right]$ the total current density calculated using the transmitted field at the graphene as the driving field, and it is assumed that the graphene sheet is on the interface between air and a substrate that has an index of refraction $n$. In this Section we take $n=1$ and independent of frequency for convenience. We take the incident field to be linearly polarized in the $x$-direction, such that $\mathbf{E}=E\hat{\mathbf{x}}$ and $\mathbf{J}=J\hat{\mathbf{x}}$. Fourier transforming these quantities, $J\left(t\right)\rightarrow\tilde{J}\left(\omega\right)$, $E^{i\left(t\right)}\left(t\right)\rightarrow\tilde{E}^{i\left(t\right)}\left(\omega\right)$ we extract the (potentially field-dependent) conductivity
\begin{equation}
\sigma\left(\omega\right)\equiv\frac{\tilde{J}\left(\omega\right)}{\tilde{E}^{t}\left(\omega\right)},
\label{eq:sigma}
\end{equation}
and thus consider the effects of incident THz field strength, phonon scattering, and neutral impurity scattering on the intraband dynamics of n-doped graphene through the lenses of the conduction band density matrix and the conductivity.

\subsection{No scattering}
To gain some intuition for the full problem, we first consider the case where there is no scattering. We set $\Gamma_{c}^{\text{out}}\left(\mathbf{k}\right)=\Gamma_{c}^{\text{in}}\left(\mathbf{k}\right)=0$, and vary $E_{\text{max}}$ from 5 kV/cm to 30 kV/cm. At low field amplitudes, we expect $\rho_{cc}\left(\mathbf{k}\right)$ to be only slightly perturbed and that the conductivity will be fully explained by a simple Drude model~\cite{Razavipour:2015}
\begin{equation}
\sigma\left(\omega\right)=\frac{2e^{2}k_{\text{B}}T\ln\left[2\cosh\left(\frac{\mu_{\text{F}}}{2k_{\text{B}}T}\right)\right]}{\pi\hbar^{2}\left(1/\tau-i\omega\right)},
\label{eq:simplesigma}
\end{equation}
where $\tau$ is the scattering time. Note that here in Subsection A, because there is no scattering, we take $\tau\rightarrow\infty$.

\begin{figure}[htbp]
\centering
\includegraphics[width=0.9\linewidth]{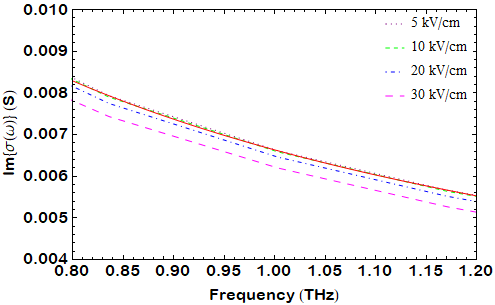}
\caption{Imaginary part of the conductivity as calculated using Eq.~\eqref{eq:sigma} when no scattering mechanisms are present for various incident field strengths. For comparison, we also plot the imaginary part of the Drude conductivity Eq.~\eqref{eq:simplesigma} when $\tau\rightarrow\infty$ as the solid red line.}
\label{fig:NSI}
\end{figure}

In  Fig.~\ref{fig:NSI} we plot the imaginary part of the conductivity as given by Eq.~\eqref{eq:sigma}. The real part of the conductivity is zero, as there is no scattering. At a field strength of only 5~kV/cm, the imaginary part of the conductivity agrees well with the Drude model with an infinite scattering time. However, as the field strength increases we leave the linear regime and the simple Drude model breaks down. Most importantly, we see that the conductivity decreases as the field amplitude increases. This modest ($\sim 6.3\%$ at 1~THz) decrease is due to the nonlinear response (clipping) that arises from the linear dispersion of graphene, as discussed in the introduction. 

\begin{figure*}[ht]
\centering
\includegraphics[width=0.6\paperwidth]{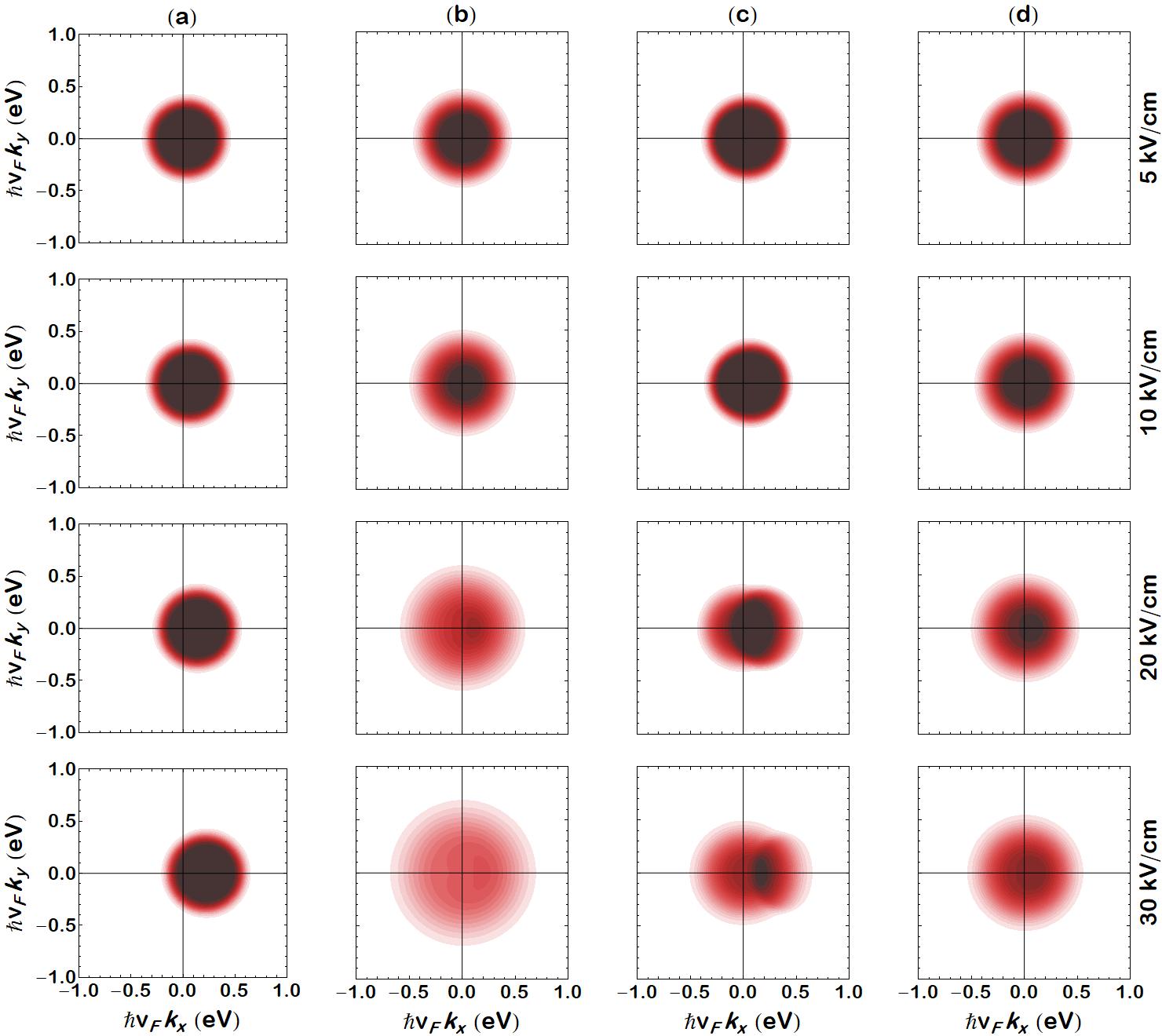}
\caption{Conduction band density matrix $\rho_{cc}\left(\mathbf{k}\right)$ at time $t=3\text{ ps}$ for various incident field amplitudes and scattering mechanisms. Black corresponds to $\rho_{cc}\left(\mathbf{k}\right)=1$ and white to $\rho_{cc}\left(\mathbf{k}\right)=0$. (a) No scattering. (b) Neutral impurity scattering only. (c) Optical phonon scattering only. (d) Both neutral impurity and optical phonon scattering.}
\label{fig:rho}
\end{figure*}

To aid in the examination of the effects of different scattering mechanisms on the nonlinear carrier dynamics, in Fig.~\ref{fig:rho} we plot the carrier density in $\mathbf{k}$-space at $t=3\text{ ps}$ for different field amplitudes and with different scattering mechanisms. We choose the time of 3~ps because, again using $\mathbf{E}^{i}\left(t\right)$ rather than $\mathbf{E}^{t}\left(t\right)$ as a first approximation, this is when the largest displacement in $\mathbf{k}$-space is expected. In Fig.~\ref{fig:rho}(a), we plot the density when there is no scattering. We see that the electrons have moved very little for an incident field of 5~kV/cm, while they have moved more than $\hbar v_{\text{F}}k_{x}=0.2\text{ eV}$ for an incident field of 30~kV/cm. With no scattering, electrons are not drawn back toward the Dirac point except by the field. As such, the current becomes saturated when the majority of electrons have a positive value of $\left|k_{x}\right|$ and have essentially maximized their velocity. This current saturation goes hand in hand with a decrease in the conductivity as the field strength is increased, as observed in Fig.~\ref{fig:NSI}, and is not explained  by a simple Drude model. However, note that this is a highly idealized system.

\subsection{Neutral impurity scattering only}
In a more realistic picture, the presence of lattice defects leads to scattering and an input scattering rate of [recall Eq.~\eqref{eq:GammaIn}]
\begin{equation}
\Gamma_{c}^{\text{in}}\left(\mathbf{k}\right)=\frac{n_{i}v_{0}^{2}}{2\pi\hbar^{2}v_{\text{F}}}\int_{0}^{2\pi}\text{d}\theta_{q}\left[1+\cos\left(\theta_{\mathbf{k},\mathbf{k}\left(\theta_{q}\right)}\right)\right]\left\vert\mathbf{k}\right\vert\rho_{cc}\left[\mathbf{k}\left(\theta_{q}\right)\right].
\label{eq:GammaImp}
\end{equation}
Following Hwang and Das Sarma~\cite{Hwang:2008} we take $n_{i}=3\times10^{10}\text{ cm}^{-2}$ and $v_{0}=1\text{ keV \AA}^{2}$ or, for a more direct comparison with $g_{\Gamma}^{2}$ and $g_{K}^{2}$ above, $n_{i}v_{0}^{2}N/A=0.573\text{ eV}^{2}$.

\begin{figure}[htbp]
\centering
\includegraphics[width=0.9\linewidth]{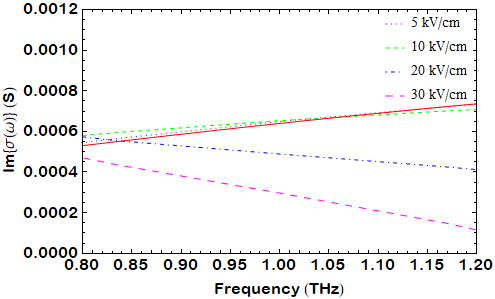}
\caption{Imaginary part of the conductivity as calculated using Eq.~\eqref{eq:sigma} when neutral impurity scattering is the only scattering mechanism, with an impurity density $n_{i}=3\times10^{10}\text{ cm}^{-2}$ and $v_{0}=1\text{ keV }\AA^{2}$, for various incident field strengths. For comparison, we also plot the imaginary part of the Drude conductivity Eq.~\eqref{eq:simplesigma} when $\tau=52\text{ fs}$ as the solid red line.}
\label{fig:II}
\end{figure}

For the same four incident field strengths as above, we plot the resulting imaginary part of the conductivity in Fig.~\ref{fig:II}. Note how it has been reduced by approximately an order of magnitude relative to Fig.~\ref{fig:NSI}, where there was no scattering. Indeed, in light of Eq.~\eqref{eq:simplesigma}, which we expect to be valid for the 5~kV/cm field, a reduction should be expected as scattering channels are introduced and $\tau$ becomes finite. To estimate the size of $\tau$ we evaluate Eq.~\eqref{eq:GammaImp} for $\left\vert\mathbf{k}\right\vert=\mu_{c}/\left(\hbar v_{\text{F}}\right)$ and $\rho_{cc}=1/2$, finding $\tau=\Gamma_{c}^{\text{in}}\left(\mathbf{k}_{\text{F}}\right)^{-1}=\Gamma_{c}^{\text{out}}\left(\mathbf{k}_{\text{F}}\right)^{-1}=52\text{ fs}$. As can be seen in Fig.~\ref{fig:II}, the Drude model of Eq.~\eqref{eq:simplesigma} with $\tau=52\text{ fs}$ agrees well with the plot for the 5~kV/cm field. However, as the field strength increases, the slope of the imaginary part of the conductivity with respect to frequency changes, and, again, the simple Drude model breaks down. 

Tracking the carriers directly in Fig.~\ref{fig:rho}(b), we see that while they are not displaced to as large a positive $k_{x}$ as in Fig.~\ref{fig:rho}(a) when there is no scattering, they are still seen to be far from equilibrium at a time of 3~ps for large field amplitudes. As neutral impurity scatterers redistribute momentum while keeping energy constant, each time carriers are driven to large $k_{x}$, they are scattered to a different point with the same $\left\vert\mathbf{k}\right\vert$. This scattering is faster at higher energies $\varepsilon_{c}\left(\mathbf{k}\right)=\hbar v_{\text{F}}\left|\mathbf{k}\right|$ because, at higher energies, there are a larger number of states to scatter into than at lower energies. As the effective scattering time $\tau=52\text{ fs}$ is much less than the period of the incident pulse (1 ps), many scattering events have occurred by the time $t=3\text{ ps}$, and so the distribution resembles a uniform disk that is displaced somewhat in the positive-$k_{x}$ direction. Moreover, the size of the electron disc at $t=3\text{ ps}$ is seen to grow with increasing field strength.

\begin{figure}[htbp]
\centering
\includegraphics[width=0.9\linewidth]{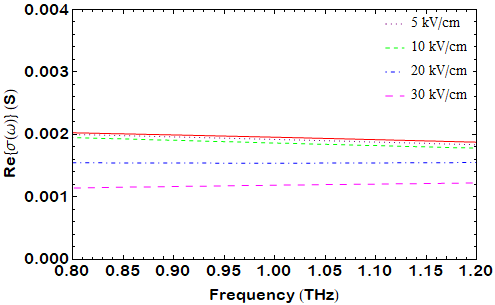}
\caption{Real part of the conductivity as calculated using Eq.~\eqref{eq:sigma} when neutral impurity scattering is the only scattering mechanism, with an impurity density $n_{i}=3\times10^{10}\text{ cm}^{-2}$ and $v_{0}=1\text{ keV }\AA^{2}$, for various incident field strengths. For comparison, we also plot the real part of the Drude conductivity Eq.~\eqref{eq:simplesigma} when $\tau=52\text{ fs}$ as the solid red line.}
\label{fig:IR}
\end{figure}

The inclusion of scattering has introduced a real part to the conductivity, which we plot in Fig.~\ref{fig:IR}. Note that for the impurity density we have chosen, the real part of the conductivity is considerably larger than the imaginary part. In comparison to the imaginary part of the conductivity (Fig.~\ref{fig:II}), it is much flatter as a function of frequency and does not change slope with increasing field strength. As in the case of the imaginary part of the conductivity, the Drude model with $\tau=52\text{ fs}$ agrees with the 5~kV/cm field plot of Fig.~\ref{fig:IR}. 

We note that the reduction of the conductivity with increasing field strength field is much more significant here than it was when there was no scattering. Indeed, here $\left|\sigma\right|$ is reduced by $\sim 39.5\%$  at 1~THz as the field strength is increased from 5~kV/cm to 30~kV/cm.  This increase in the nonlinearity of the response can largely be seen as arising from the energy dependence of the scattering.  When the field is stronger, the electrons are driven to higher energy, where the number of available scattering states is larger and hence the scattering time is shorter. 
\subsection{Phonon scattering only}
We now consider the effect of phonon scattering alone, i.e. we set
\begin{align}
&\Gamma_{c}^{\text{in}}\left(\mathbf{k}\right)\nonumber\\
&=\frac{A}{2\pi\hbar^{2}v_{\text{F}}N}\int_{0}^{2\pi}\text{d}\theta_{q}\nonumber\\
&\times\left\{ 2g_{\Gamma}^{2}\left(n_{\Gamma}+1\right)k_{\Gamma}^{+}\rho_{cc}\left[\mathbf{k}_{\Gamma}^{+}\left(\theta_{q}\right)\right]\right.\nonumber\\
&\quad+2g_{\Gamma}^{2}n_{\Gamma}k_{\Gamma}^{-}\rho_{cc}\left[\mathbf{k}_{\Gamma}^{-}\left(\theta_{q}\right)\right]\nonumber\\
&\quad+g_{K}^{2}\left[1-\cos\left(\theta_{\mathbf{k},\mathbf{k}_{K}^{+}\left(\theta_{q}\right)}\right)\right]\left(n_{K}+1\right)k_{K}^{+}\rho_{cc}\left[\mathbf{k}_{K}^{+}\left(\theta_{q}\right)\right]\nonumber\\
&\quad+\left.g_{K}^{2}\left[1-\cos\left(\theta_{\mathbf{k},\mathbf{k}_{K}^{-}\left(\theta_{q}\right)}\right)\right]n_{K}k_{K}^{-}\rho_{cc}\left[\mathbf{k}_{K}^{-}\left(\theta_{q}\right)\right]\right\}.
\end{align}
This mechanism of scattering differs from impurity scattering in several important respects. First, there are two distinct modes of scattering: phonon absorption and phonon emission. In contrast to impurity scattering, both of these mechanisms are \textit{inelastic}. In the first, electrons are scattered to higher energy states in the conduction band due to the absorption of an optical phonon, with the rate being proportional to the optical phonon population. In the second, electrons that have enough energy stimulate the emission of a phonon and scatter to an empty state with lower energy. We therefore expect the effects of phonon scattering to be very different than impurity scattering and, in particular, we expect phonon emission scattering to exhibit an energy threshold and therefore strongly depend on the THz field amplitude.

\begin{figure}[htbp]
\centering
\includegraphics[width=0.9\linewidth]{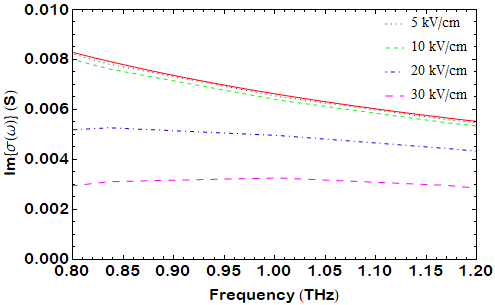}
\caption{Imaginary part of the conductivity as calculated using Eq.~\eqref{eq:sigma} when optical phonon scattering is the only scattering mechanism, for various incident field strengths. For comparison, we also plot the imaginary part of the Drude conductivity Eq.~\eqref{eq:simplesigma} when $\tau\rightarrow\infty$ as the solid red line.}
\label{fig:PI}
\end{figure}

In Fig.~\ref{fig:PI} we plot the imaginary part of the conductivity resulting from our simulations involving optical phonon scattering as the only scattering channels. Note that for a field strength of only 5~kV/cm, again the curve closely matches that produced by the simple Drude model with an infinite scattering time. In contrast to Fig.~\ref{fig:II}, where neutral impurity scattering is always present, here we see that, at low field strengths, the carrier response is essentially identical to the response when no scattering channels are present. This is because, for low field amplitudes, all of the electron states with energies $\hbar\omega_{\Gamma\left(K\right)}$ below occupied states are also occupied, and so phonon emission is essentially forbidden and only phonon absorption is possible. Additionally, at room temperature the phonon populations are almost negligible ($n_{\Gamma}=0.00051$, $n_{K}=0.0021$) and so scattering via phonon absorption is also very weak. However, by the same token, the dependence of $\text{Im}\left(\sigma\right)$ on field strength is much stronger than for neutral impurities, because phonon emission only becomes possible when carriers are driven to high enough energies $\varepsilon_{c}\left(\mathbf{k}\right)=\hbar v_{\text{F}}\left\vert\mathbf{k}\right\vert$ such that there are unoccupied states with energies $\hbar\omega_{\Gamma\left(K\right)}$ below occupied states.

Tracking the carriers in Fig.~\ref{fig:rho}(c), we see that, compared to Fig.~\ref{fig:rho}(b) where neutral impurity scattering is the only scattering mechanism, there are more carriers near to the Dirac point at all field amplitudes. This is because when phonon scattering is present, once carriers are driven to large enough $\varepsilon_{c}\left(\mathbf{k}\right)=\hbar v_{\text{F}}\left\vert\mathbf{k}\right\vert$, they can create a phonon and give up energy to move nearer to the Dirac point. We see that for the higher field amplitudes that there is a high-density region, with a rather well-defined edge relatively close to the Dirac point, but displaced in the positive $k_{x}$ direction.  This feature arises from the phonon emission process and does not dissipate quickly, because here there is no elastic scattering to redistribute the carriers.

\begin{figure}[htbp]
\centering
\includegraphics[width=0.9\linewidth]{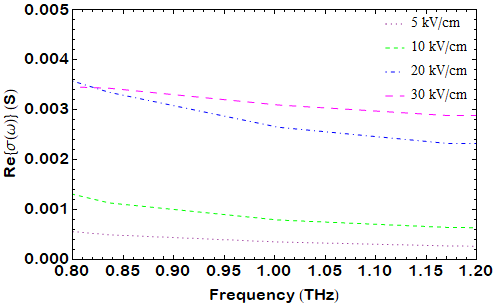}
\caption{Real part of the conductivity as calculated using Eq.~\eqref{eq:sigma} when optical phonon scattering is the only scattering mechanism, for various incident field strengths.}
\label{fig:PR}
\end{figure}

In Fig.~\ref{fig:PR} we plot the real part of the conductivity. In contrast to what was found for neutral impurities, the real part of the conductivity is of similar size or smaller than the imaginary part. In addition, the real part of the conductivity increases by almost an order of magnitude when the field is increased from 5~kV/cm to 30~kV/cm. Both of these effects are due to the absorptive nature of photon scattering when there is phonon emission. Looking at  Fig.~\ref{fig:rho}(c), the threshold for significant phonon emission is seen to occur between field strengths of 20~kV/cm and 30~kV/cm. Comparing with Fig.~\ref{fig:rho}(a), this is seen to coincide with field strengths for which the electron disc is driven to energies larger than $\hbar\omega_{\Gamma\left(K\right)}$ in the absence of any scattering (in particular, the average values of $\hbar v_{\text{F}}k_{x}$ in Fig.~\ref{fig:rho}(a) are 0.136~eV and 0.219~eV for the 20~kV/cm and 30~kV/cm fields, respectively).

Again, we note that the field-induced decrease in the conductivity is larger here than when there was no scattering. It is, however, similar to when there was neutral impurity scattering only. Here $\left|\sigma\right|$ is reduced by $\sim 31.2\%$ at 1~THz when the field is increased from 5~kV/cm to 30~kV/cm. This large change is due to the sudden turn-on of phonon-emission scattering when the field reaches a threshold value.

To summarize, although neutral impurity scattering is significant for all field strengths and serves to redistribute carrier momentum about $\mathbf{k}=0$, phonon scattering is only significant for moderate to high field strengths and serves to drive carriers to lower energies. The effect of neutral impurity scattering on the conductivity at low field strengths can be explained with a simple Drude model, while, at low field strengths, the effect of phonon scattering is negligible. As the field strength is increased, the simple Drude model breaks down, and the imaginary part of the conductivity decreases, in line with our intuition that the effects of scattering are increasing. Indeed, the scattering-out rates for each of the scattering mechanisms considered depend on the number of states available to scatter into [recall Eq.~\eqref{eq:GammaIn}]. As these rates are proportional to $2\pi\left\vert\mathbf{q}\right\vert$, where $\mathbf{q}$ is the final momentum, it is expected that they should increase as carriers are driven to larger $\hbar v_{\text{F}} \left\vert\mathbf{k}\right\vert$ and the number of states with the required final momentum increases.

\subsection{Neutral impurity and phonon scattering}
\begin{figure}[htbp]
\centering
\includegraphics[width=0.9\linewidth]{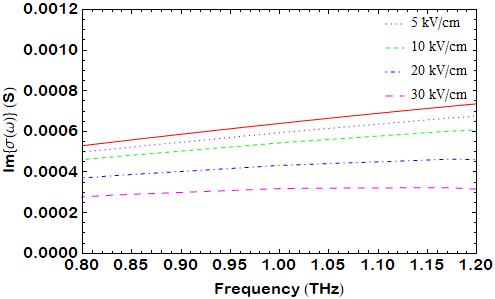}
\caption{Imaginary part of the conductivity as calculated using Eq.~\eqref{eq:sigma} when both neutral impurity scattering and phonon scattering are present, with an impurity density $n_{i}=3\times10^{10}\text{ cm}^{-2}$ and $v_{0}=1\text{ keV }\AA^{2}$, for various incident field strengths. For comparison, we also plot the imaginary part of the Drude conductivity Eq.~\eqref{eq:simplesigma} when $\tau=52\text{ fs}$ as the solid red line.}
\label{fig:IPI}
\end{figure}

\begin{figure}[htbp]
\centering
\includegraphics[width=0.9\linewidth]{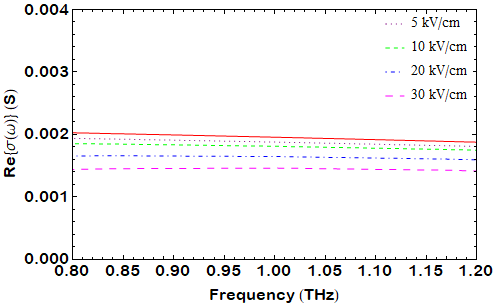}
\caption{Real part of the conductivity as calculated using Eq.~\eqref{eq:sigma} when both neutral impurity scattering and phonon scattering are present, with an impurity density $n_{i}=3\times10^{10}\text{ cm}^{-2}$ and $v_{0}=1\text{ keV }\AA^{2}$, for various incident field strengths. For comparison, we also plot the real part of the Drude conductivity Eq.~\eqref{eq:simplesigma} when $\tau=52\text{ fs}$ as the solid red line.}
\label{fig:IPR}
\end{figure}

In this subsection we include both neutral impurity and phonon scattering, again using $n_{i}=3\times10^{10}\text{ cm}^{-2}$ and $v_{0}=1\text{ keV }\AA^{2}$. We plot the resulting imaginary and real parts of the conductivity in Figs.~\ref{fig:IPI} and~\ref{fig:IPR}, respectively. The imaginary part of the conductivity has become flatter as a function of frequency relative to Fig.~\ref{fig:II}, where only neutral impurity scattering was included. Compared to Fig.~\ref{fig:PR}, where only phonon scattering is present, the real part of the conductivity has also become flatter as a function of frequency. We note that although the field-induced change in the conductivity is still very large, it is somewhat more modest here than when there is only photon or only neutral impurity scattering, with $\left|\sigma\right|$ being reduced by about 24.0\% at 1~THz when the field is increased from 5~kV/cm to 30~kV/cm. In addition, the sudden onset of the nonlinear response  that was present when there was only phonon scattering is greatly softened here due to the redistribution of carriers by neutral impurity scattering. At low field strengths, there is still rather good agreement with the simple Drude model with a scattering time of $\tau=52\text{ fs}$, for at low field strengths phonon scattering plays essentially no role. We believe the small discrepancy to be due to phonon emission enabled by the redistribution of carriers by neutral impurity scattering, as seen by comparing Fig.~\ref{fig:rho}(c) to Fig.~\ref{fig:rho}(d).

Looking at Fig.~\ref{fig:rho}(d) more closely, we see that the distribution of carriers has inherited characteristics from both scattering mechanisms at all field strengths. In particular, carriers are more symmetric about $\mathbf{k}=0$ than when no neutral impurity scattering is present, and nearer the Dirac point than when no phonon scattering is present. In addition, the sharp edge in the density that was seen at high fields when there was only phonon scattering is now gone, as a result of the redistribution in $\mathbf{k}$-space due to the neutral impurity scattering

\subsection{Semi-empirical scattering}
\begin{figure}[htbp]
\centering
\includegraphics[width=0.9\linewidth]{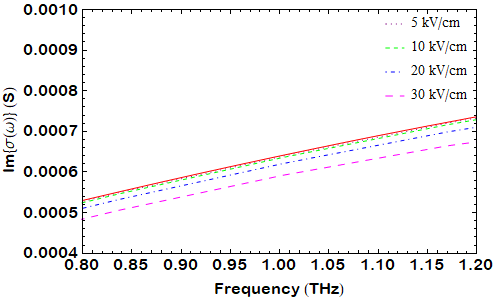}
\caption{Imaginary part of the conductivity as calculated using Eq.~\eqref{eq:SEM} with a relaxation time of $\tau_{c}=52\text{ fs}$, for various incident field strengths. For comparison, we also plot the imaginary part of the Drude conductivity Eq.~\eqref{eq:simplesigma} when $\tau=52\text{ fs}$ as the solid red line.}
\label{fig:SEI}
\end{figure}

\begin{figure}[htbp]
\centering
\includegraphics[width=0.9\linewidth]{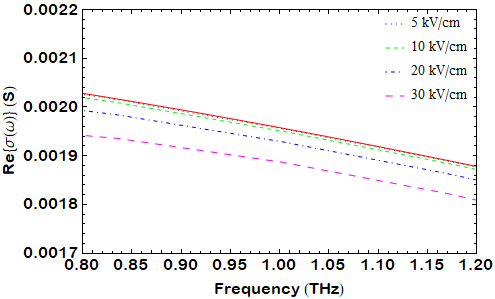}
\caption{Real part of the conductivity as calculated using Eq.~\eqref{eq:SEM} with a relaxation time of $\tau_{c}=52\text{ fs}$, for various incident field strengths. For comparison, we also plot the real part of the Drude conductivity Eq.~\eqref{eq:simplesigma} when $\tau=52\text{ fs}$ as the solid red line.}
\label{fig:SER}
\end{figure}

Finally, we compare the results above with an earlier semi-empirical model of scattering dynamics~\cite{Al-Naib:2014,Al-Naib:2015}
\begin{equation}
\frac{\text{d}\rho_{cc}\left(\mathbf{k}\right)}{\text{d}t}=-\frac{e\mathbf{E}\left(t\right)}{\hbar}\cdot\nabla_{\mathbf{k}}\rho_{cc}\left(\mathbf{k}\right)-\frac{\rho_{cc}\left(\mathbf{k}\right)-\rho_{cc}^{\text{init}}\left(\mathbf{k}\right)}{\tau_{c}},
\label{eq:SEM}
\end{equation}
where $\rho_{cc}^{\text{init}}\left(\mathbf{k}\right)$ is the thermal state of Eq.~\eqref{eq:rhot} and $\tau_{c}$ a phenomenological relaxation time. Using $\tau_{c}=52\text{ fs}$, we plot the resulting imaginary and real parts of the conductivity in Fig.~\ref{fig:SEI} and Fig.~\ref{fig:SER}, respectively. 

Note that, here, $\left|\sigma\right|$ only decreases by $\sim 3.9\%$ at 1~THz when the field is increased from 5 to 30 kV/cm. This is less than in any of the other scattering scenarios. It is weaker than when there is no scattering, because the scattering limits how far in $\mathbf{k}$-space the electrons are driven from the Dirac point, which thus reduces the effect of the intrinsic nonlinearity. The field-dependence is less than when the microscopic scattering is included because, unlike in Eq.~\eqref{eq:SBE}, in Eq.~\eqref{eq:SEM} there is no mechanism for carriers at different $\mathbf{k}$ to experience different scattering rates---they all experience the same rate. This causes carriers at large $\left\vert\mathbf{k}\right\vert$ to experience too small a scattering rate and carriers at small $\left\vert\mathbf{k}\right\vert$ to experience too large a scattering rate. While this may be a reasonable approximation at weak field strengths, where the diferences in scattering rates experienced by carriers at each $\mathbf{k}$ are small, it cannot capture the effect of neutral impurity scattering, nor neutral impurity plus optical phonon scattering, as carriers are driven to larger $\mathbf{k}$ by stronger fields and these differences become larger. We note that one could improve this model somewhat by allowing $\tau_{c}$ to be energy dependent, but it would still not be able to capture effects such as phonon emission rates that depend critically on the availability of states into which the carriers can scatter.

\section{Comparison to experiment}
\label{sec:Application}
Having examined the carrier dynamics and field-dependent conductivities for the five different scattering scenarios above, we now examine whether our model can at least qualitatively explain a recent THz TDS experiment, in which transmission through monolayer graphene was shown to increase with increasing field strength~\cite{Hafez:2014}.

In Ref.~\onlinecite{Hafez:2014}, Hafez et al.~measured the transmission of intense THz pulses through an n-doped graphene sheet with a carrier density of  $n_{c}=7.62\times 10^{12}\text{ cm}^{-2}$ on a silicon carbide (SiC) substrate and examined it relative to the transmission through the SiC substrate alone. At the central frequency of 0.7~THz, they observed an approximately 4.3\% increase in this ratio, known as the (normalized) transmission, when increasing peak field strength $E_\text{max}$ from 13~kV/cm to 63~kV/cm. In our notation, this transmission spectrum is given by
\begin{equation}
\text{Transmission}=\frac{n+1}{2}\left\vert \frac{\tilde{E}^{t}\left(\omega\right)}{\tilde{E}^{i}\left(\omega\right)}\right\vert\equiv\text{Tr}.
\label{eq:Tr}
\end{equation}
Hafez et al.~used  the semi-empirical model of Eq.~\eqref{eq:SEM} to attempt to explain their data, and found that it required using $\tau_{c}$ as a fitting parameter, allowing $\tau_{c}$ to vary with field strength. For completeness, here we first try to model their results using Eq.~\eqref{eq:SEM} before applying our full model including neutral impurity and optical phonon scattering.

We take the pulses to be characterized by $f=0.7\text{ THz}$, and scale the pulse duration $\Delta_{t}=1.0/0.7=1.43\text{ ps}$ and temporal offset $t_{0}=3.0/0.7=4.3\text{ ps}$, running simulations from $t=0\text{ ps}$ to $t=8.6\text{ ps}$ with 
normalization constant $N_{E}=0.859166$, so that the pulse shape remains the same as in the above scenarios. We also increase the gridpoint density~\cite{Grid} to $901\times 901$ points. 

\begin{figure}[htbp]
\centering
\includegraphics[width=0.9\linewidth]{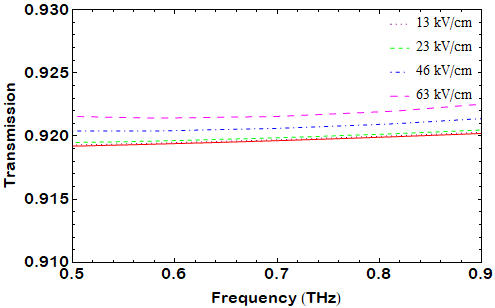}
\caption{Transmission as obtained from Eq.~\eqref{eq:Tr} using the semiempirical model of Eq.~\eqref{eq:SEM} with a relaxation time of $\tau_{c}=25.5\text{ fs}$, for various incident field strengths. For comparison, we also plot the transmission  as calculated using the Drude model of Eq.~\eqref{eq:simplesigma} when $\tau=25.5\text{ fs}$ as the solid red line.}
\label{fig:SE3}
\end{figure}

We first model the transmission spectrum using the semi-empirical model of Eq.~\eqref{eq:SEM}. We take the refractive index of SiC at THz frequencies to be $n=3$, and treat the scattering time $\tau_{c}$ as a free parameter. We adjust $\tau_{c}$ to match the experimental transmission measured in Ref.~\onlinecite{Hafez:2014} at the central frequency of 0.7 THz for a field amplitude of 13~kV/ cm. This yields $\tau_{c}=25.5\text{ fs}$, which we use here for all four measured field amplitudes. In Fig.~\ref{fig:SE3}, we plot the resulting transmission as the field strength is increased from 13~kV/cm to 63~kV/cm. Note that it does not change much with increasing field strength ($\sim~0.2\%$ from 13~kV/cm to 63~kV/cm at 0.7~THz), whereas in Hafez et al.~the transmission ratio at 0.7~THz was observed to increase from $\sim~0.92$ at a field of 13~kV/cm to $\sim~0.96$  for a field of 63~kV/cm. Indeed $\tau_{c}$ would have to be adjusted with $E_\text{max}$ to maintain agreement with the experimental data of Ref.~\onlinecite{Hafez:2014}.

For simulations with microscopic scattering included, we first estimate the neutral impurity density $n_{i}$. At a peak field strength of only 13~kV/cm, we expect that optical phonon scattering will be essentially negligible and neutral impurity scattering will be the dominant scattering mechanism, and so begin by calculating the $n_{i}$ required for $\Gamma_{c}^{\text{in}}\left(\mathbf{k}_{\text{F}}\right)^{-1}=\Gamma_{c}^{\text{out}}\left(\mathbf{k}_{\text{F}}\right)^{-1}=25.5\text{ fs}$. Again taking $\left\vert\mathbf{k}\right\vert=\mu_{c}/\left(\hbar v_{\text{F}}\right)$ in Eq.~\eqref{eq:GammaImp}, along with $\rho_{cc}=1/2$, we find $n_{i}=6.9\times 10^{10}\text{ cm}^{-2}$. While this forms an initial point from which to search for the appropriate neutral impurity density, we ultimately obtain slightly better agreement with the experimentally-obtained transmission at 0.7~THz and a field strength of 13~kV/cm using $n_{i}=6.6\times 10^{10}\text{ cm}^{-2}$ in our full model.

\begin{figure}[htbp]
\centering
\includegraphics[width=0.9\linewidth]{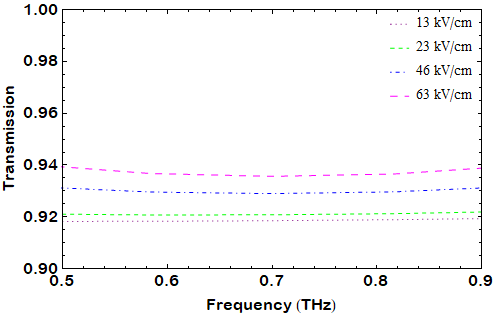}
\caption{Transmission as calculated using Eq.~\eqref{eq:Tr} including the effects of both neutral impurity and optical phonon scattering, for various incident field strengths.}
\label{fig:TA}
\end{figure}

In Fig.~\ref{fig:TA} we plot the transmission as calculated using the full microscopic model of Eq.~\eqref{eq:SBE}. As in Ref.~\onlinecite{Hafez:2014}, the transmission is seen to increase with increasing field strength. The increase in the transmission at 0.7~THz as the field strength is increased from 13~kV/cm to 63~kV/cm here is approximately 1.9\%, which is in qualitative agreement with the experimental result of Ref.~\onlinecite{Hafez:2014}, and is much closer to the experimental results than the semi-empirical model of Eq.~\eqref{eq:SEM}. We note that our field is not identical to the field used in the experiment, and this may be part of the source of the discrepancy between the transmission seen in experiment and in our simulations.

\section{Conclusion}
\label{sec:Conclusion}
In conclusion, we have presented a microscopic theory of carrier scattering in graphene, and used it to simulate carrier dynamics as well as conductivities as functions of field strength. In the absence of scattering, as field strength is increased, current clipping becomes apparent and, although the nonlinearity is rather modest for the field strengths considered, a simple Drude model cannot explain the resulting conductivity. When only neutral impurity scattering is present the resulting conductivity is reduced and is affected much more by the field amplitude than when no scattering is present. When only phonon scattering is present, again there is a strong dependence on the field amplitude but, in contrast to the case of neutral impurity scattering, there is a threshold field strength required before the resulting conductivity differs much from the simple Drude model. Finally, when both scattering mechanisms are present, the conductivity still changes very significantly as the field strength increases, but by somewhat less than when only one of the two scattering mechanisms is present. We note that at low fields $\le 5\text{ kV/cm}$, the simple Drude model with a phenomenological scattering time gives good agreement with our microscopic scattering models.  However, it fails quite dramatically at higher field amplitudes.  This is very important for experiments and devices relying on the THz response of graphene, because it is routine now to attain THz fields that are well above 5~kV/cm.

We have compared our simulation of the field-dependent transmission of a THz pulse through the graphene with both neutral impurity and optical phonon scattering with recent experimental results and find qualitative agreement. Although this agreement is encouraging, it appears that either the physical parameters we have used are not quite what are found in the actual system or there may still be elements missing from our model. In particular, the exact results of our model will depend quite strongly on the optical phonon energies and the coupling constants and these are not exactly known and were not used as fitting parameters in our simulations. Additionally, the gap between theory and experiment could potentially be further reduced by including phonon dynamics and/or other scattering mechanisms, such as charged impurities and carrier-carrier scattering, both of which we plan to turn to in future work. To further such work, it will be important that other experimental groups repeat these experiments for different graphene samples, substrates and pulse shapes to better understand the various dependencies.

\bibliography{scattering}

%merlin.mbs apsrev4-1.bst 2010-07-25 4.21a (PWD, AO, DPC) hacked
%Control: key (0)
%Control: author (0) dotless jnrlst
%Control: editor formatted (1) identically to author
%Control: production of article title (0) allowed
%Control: page (1) range
%Control: year (0) verbatim
%Control: production of eprint (0) enabled
\begin{thebibliography}{23}%
\makeatletter
\providecommand \@ifxundefined [1]{%
 \@ifx{#1\undefined}
}%
\providecommand \@ifnum [1]{%
 \ifnum #1\expandafter \@firstoftwo
 \else \expandafter \@secondoftwo
 \fi
}%
\providecommand \@ifx [1]{%
 \ifx #1\expandafter \@firstoftwo
 \else \expandafter \@secondoftwo
 \fi
}%
\providecommand \natexlab [1]{#1}%
\providecommand \enquote  [1]{``#1''}%
\providecommand \bibnamefont  [1]{#1}%
\providecommand \bibfnamefont [1]{#1}%
\providecommand \citenamefont [1]{#1}%
\providecommand \href@noop [0]{\@secondoftwo}%
\providecommand \href [0]{\begingroup \@sanitize@url \@href}%
\providecommand \@href[1]{\@@startlink{#1}\@@href}%
\providecommand \@@href[1]{\endgroup#1\@@endlink}%
\providecommand \@sanitize@url [0]{\catcode `\\12\catcode `\$12\catcode
  `\&12\catcode `\#12\catcode `\^12\catcode `\_12\catcode `\%12\relax}%
\providecommand \@@startlink[1]{}%
\providecommand \@@endlink[0]{}%
\providecommand \url  [0]{\begingroup\@sanitize@url \@url }%
\providecommand \@url [1]{\endgroup\@href {#1}{\urlprefix }}%
\providecommand \urlprefix  [0]{URL }%
\providecommand \Eprint [0]{\href }%
\providecommand \doibase [0]{http://dx.doi.org/}%
\providecommand \selectlanguage [0]{\@gobble}%
\providecommand \bibinfo  [0]{\@secondoftwo}%
\providecommand \bibfield  [0]{\@secondoftwo}%
\providecommand \translation [1]{[#1]}%
\providecommand \BibitemOpen [0]{}%
\providecommand \bibitemStop [0]{}%
\providecommand \bibitemNoStop [0]{.\EOS\space}%
\providecommand \EOS [0]{\spacefactor3000\relax}%
\providecommand \BibitemShut  [1]{\csname bibitem#1\endcsname}%
\let\auto@bib@innerbib\@empty
%</preamble>
\bibitem [{\citenamefont {Novoselov}\ \emph {et~al.}(2005)\citenamefont
  {Novoselov}, \citenamefont {Jiang}, \citenamefont {Schedin}, \citenamefont
  {Booth}, \citenamefont {Khotkevich}, \citenamefont {Morozov},\ and\
  \citenamefont {Geim}}]{Novoselov:2004}%
  \BibitemOpen
  \bibfield  {author} {\bibinfo {author} {\bibfnamefont {K.~S.}\ \bibnamefont
  {Novoselov}}, \bibinfo {author} {\bibfnamefont {D.}~\bibnamefont {Jiang}},
  \bibinfo {author} {\bibfnamefont {F.}~\bibnamefont {Schedin}}, \bibinfo
  {author} {\bibfnamefont {T.~J.}\ \bibnamefont {Booth}}, \bibinfo {author}
  {\bibfnamefont {V.~V.}\ \bibnamefont {Khotkevich}}, \bibinfo {author}
  {\bibfnamefont {S.~V.}\ \bibnamefont {Morozov}}, \ and\ \bibinfo {author}
  {\bibfnamefont {A.~K.}\ \bibnamefont {Geim}},\ }\bibfield  {title} {\enquote
  {\bibinfo {title} {Two-dimensional atomic crystals},}\ }\href {\doibase
  10.1073/pnas.0502848102} {\bibfield  {journal} {\bibinfo  {journal} {Proc.
  Natl. Acad. Sci. U.S.A.}\ }\textbf {\bibinfo {volume} {102}},\ \bibinfo
  {pages} {10451--10453} (\bibinfo {year} {2005})}\BibitemShut {NoStop}%
\bibitem [{\citenamefont {Randviir}\ \emph {et~al.}(2014)\citenamefont
  {Randviir}, \citenamefont {Brownson},\ and\ \citenamefont
  {Banks}}]{Randvir:2014}%
  \BibitemOpen
  \bibfield  {author} {\bibinfo {author} {\bibfnamefont {Edward~P.}\
  \bibnamefont {Randviir}}, \bibinfo {author} {\bibfnamefont {Dale~A.C.}\
  \bibnamefont {Brownson}}, \ and\ \bibinfo {author} {\bibfnamefont {Craig~E.}\
  \bibnamefont {Banks}},\ }\bibfield  {title} {\enquote {\bibinfo {title} {A
  decade of graphene research: production, applications and outlook},}\ }\href
  {\doibase https://doi.org/10.1016/j.mattod.2014.06.001} {\bibfield  {journal}
  {\bibinfo  {journal} {Mater. Today}\ }\textbf {\bibinfo {volume} {17}},\
  \bibinfo {pages} {426 -- 432} (\bibinfo {year} {2014})}\BibitemShut {NoStop}%
\bibitem [{\citenamefont {Zhu}\ \emph {et~al.}(2018)\citenamefont {Zhu},
  \citenamefont {Ji}, \citenamefont {Cheng},\ and\ \citenamefont
  {Ruoff}}]{Zhu:2018}%
  \BibitemOpen
  \bibfield  {author} {\bibinfo {author} {\bibfnamefont {Yanwu}\ \bibnamefont
  {Zhu}}, \bibinfo {author} {\bibfnamefont {Hengxing}\ \bibnamefont {Ji}},
  \bibinfo {author} {\bibfnamefont {Hui-Ming}\ \bibnamefont {Cheng}}, \ and\
  \bibinfo {author} {\bibfnamefont {Rodney~S}\ \bibnamefont {Ruoff}},\
  }\bibfield  {title} {\enquote {\bibinfo {title} {Mass production and
  industrial applications of graphene materials},}\ }\href {\doibase
  10.1093/nsr/nwx055} {\bibfield  {journal} {\bibinfo  {journal} {Natl. Sci.
  Rev.}\ }\textbf {\bibinfo {volume} {5}},\ \bibinfo {pages} {90--101}
  (\bibinfo {year} {2018})}\BibitemShut {NoStop}%
\bibitem [{\citenamefont {Maeng}\ \emph {et~al.}(2012)\citenamefont {Maeng},
  \citenamefont {Lim}, \citenamefont {Chae}, \citenamefont {Lee}, \citenamefont
  {Choi},\ and\ \citenamefont {Son}}]{Maeng:2012}%
  \BibitemOpen
  \bibfield  {author} {\bibinfo {author} {\bibfnamefont {Inhee}\ \bibnamefont
  {Maeng}}, \bibinfo {author} {\bibfnamefont {Seongchu}\ \bibnamefont {Lim}},
  \bibinfo {author} {\bibfnamefont {Seung~Jin}\ \bibnamefont {Chae}}, \bibinfo
  {author} {\bibfnamefont {Young~Hee}\ \bibnamefont {Lee}}, \bibinfo {author}
  {\bibfnamefont {Hyunyong}\ \bibnamefont {Choi}}, \ and\ \bibinfo {author}
  {\bibfnamefont {Joo-Hiuk}\ \bibnamefont {Son}},\ }\bibfield  {title}
  {\enquote {\bibinfo {title} {Gate-controlled nonlinear conductivity of dirac
  fermion in graphene field-effect transistors measured by terahertz
  time-domain spectroscopy},}\ }\href {\doibase 10.1021/nl202442b} {\bibfield
  {journal} {\bibinfo  {journal} {Nano Lett.}\ }\textbf {\bibinfo {volume}
  {12}},\ \bibinfo {pages} {551--555} (\bibinfo {year} {2012})}\BibitemShut
  {NoStop}%
\bibitem [{\citenamefont {Ren}\ \emph {et~al.}(2012)\citenamefont {Ren},
  \citenamefont {Zhang}, \citenamefont {Yao}, \citenamefont {Sun},
  \citenamefont {Kaneko}, \citenamefont {Yan}, \citenamefont {Nanot},
  \citenamefont {Jin}, \citenamefont {Kawayama}, \citenamefont {Tonouchi},
  \citenamefont {Tour},\ and\ \citenamefont {Kono}}]{Ren:2012}%
  \BibitemOpen
  \bibfield  {author} {\bibinfo {author} {\bibfnamefont {Lei}\ \bibnamefont
  {Ren}}, \bibinfo {author} {\bibfnamefont {Qi}~\bibnamefont {Zhang}}, \bibinfo
  {author} {\bibfnamefont {Jun}\ \bibnamefont {Yao}}, \bibinfo {author}
  {\bibfnamefont {Zhengzong}\ \bibnamefont {Sun}}, \bibinfo {author}
  {\bibfnamefont {Ryosuke}\ \bibnamefont {Kaneko}}, \bibinfo {author}
  {\bibfnamefont {Zheng}\ \bibnamefont {Yan}}, \bibinfo {author} {\bibfnamefont
  {Sébastien}\ \bibnamefont {Nanot}}, \bibinfo {author} {\bibfnamefont
  {Zhong}\ \bibnamefont {Jin}}, \bibinfo {author} {\bibfnamefont {Iwao}\
  \bibnamefont {Kawayama}}, \bibinfo {author} {\bibfnamefont {Masayoshi}\
  \bibnamefont {Tonouchi}}, \bibinfo {author} {\bibfnamefont {James~M.}\
  \bibnamefont {Tour}}, \ and\ \bibinfo {author} {\bibfnamefont {Junichiro}\
  \bibnamefont {Kono}},\ }\bibfield  {title} {\enquote {\bibinfo {title}
  {Terahertz and infrared spectroscopy of gated large-area graphene},}\ }\href
  {\doibase 10.1021/nl301496r} {\bibfield  {journal} {\bibinfo  {journal} {Nano
  Lett.}\ }\textbf {\bibinfo {volume} {12}},\ \bibinfo {pages} {3711--3715}
  (\bibinfo {year} {2012})},\ \bibinfo {note} {pMID: 22663563}\BibitemShut
  {NoStop}%
\bibitem [{\citenamefont {Paul}\ \emph {et~al.}(2013)\citenamefont {Paul},
  \citenamefont {Chang}, \citenamefont {Thompson}, \citenamefont {Stickel},
  \citenamefont {Wardini}, \citenamefont {Choi}, \citenamefont {Minot},
  \citenamefont {Hou}, \citenamefont {Nees}, \citenamefont {Norris},\ and\
  \citenamefont {Lee}}]{Paul:2013}%
  \BibitemOpen
  \bibfield  {author} {\bibinfo {author} {\bibfnamefont {M~J}\ \bibnamefont
  {Paul}}, \bibinfo {author} {\bibfnamefont {Y~C}\ \bibnamefont {Chang}},
  \bibinfo {author} {\bibfnamefont {Z~J}\ \bibnamefont {Thompson}}, \bibinfo
  {author} {\bibfnamefont {A}~\bibnamefont {Stickel}}, \bibinfo {author}
  {\bibfnamefont {J}~\bibnamefont {Wardini}}, \bibinfo {author} {\bibfnamefont
  {H}~\bibnamefont {Choi}}, \bibinfo {author} {\bibfnamefont {E~D}\
  \bibnamefont {Minot}}, \bibinfo {author} {\bibfnamefont {B}~\bibnamefont
  {Hou}}, \bibinfo {author} {\bibfnamefont {J~A}\ \bibnamefont {Nees}},
  \bibinfo {author} {\bibfnamefont {T~B}\ \bibnamefont {Norris}}, \ and\
  \bibinfo {author} {\bibfnamefont {Yun-Shik}\ \bibnamefont {Lee}},\ }\bibfield
   {title} {\enquote {\bibinfo {title} {High-field terahertz response of
  graphene},}\ }\href {http://stacks.iop.org/1367-2630/15/i=8/a=085019}
  {\bibfield  {journal} {\bibinfo  {journal} {New J. Phys.}\ }\textbf {\bibinfo
  {volume} {15}},\ \bibinfo {pages} {085019} (\bibinfo {year}
  {2013})}\BibitemShut {NoStop}%
\bibitem [{\citenamefont {Hafez}\ \emph {et~al.}(2014)\citenamefont {Hafez},
  \citenamefont {Al-Naib}, \citenamefont {Oguri}, \citenamefont {Sekine},
  \citenamefont {Dignam}, \citenamefont {Ibrahim}, \citenamefont {Cooke},
  \citenamefont {Tanaka}, \citenamefont {Komori}, \citenamefont {Hibino},\ and\
  \citenamefont {Ozaki}}]{Hafez:2014}%
  \BibitemOpen
  \bibfield  {author} {\bibinfo {author} {\bibfnamefont {H.~A.}\ \bibnamefont
  {Hafez}}, \bibinfo {author} {\bibfnamefont {I.}~\bibnamefont {Al-Naib}},
  \bibinfo {author} {\bibfnamefont {K.}~\bibnamefont {Oguri}}, \bibinfo
  {author} {\bibfnamefont {Y.}~\bibnamefont {Sekine}}, \bibinfo {author}
  {\bibfnamefont {M.~M.}\ \bibnamefont {Dignam}}, \bibinfo {author}
  {\bibfnamefont {A.}~\bibnamefont {Ibrahim}}, \bibinfo {author} {\bibfnamefont
  {D.~G.}\ \bibnamefont {Cooke}}, \bibinfo {author} {\bibfnamefont
  {S.}~\bibnamefont {Tanaka}}, \bibinfo {author} {\bibfnamefont
  {F.}~\bibnamefont {Komori}}, \bibinfo {author} {\bibfnamefont
  {H.}~\bibnamefont {Hibino}}, \ and\ \bibinfo {author} {\bibfnamefont
  {T.}~\bibnamefont {Ozaki}},\ }\bibfield  {title} {\enquote {\bibinfo {title}
  {Nonlinear transmission of an intense terahertz field through monolayer
  graphene},}\ }\href {\doibase 10.1063/1.4902096} {\bibfield  {journal}
  {\bibinfo  {journal} {AIP Adv.}\ }\textbf {\bibinfo {volume} {4}},\ \bibinfo
  {pages} {117118} (\bibinfo {year} {2014})}\BibitemShut {NoStop}%
\bibitem [{\citenamefont {Tomadin}\ \emph {et~al.}(2018)\citenamefont
  {Tomadin}, \citenamefont {Hornett}, \citenamefont {Wang}, \citenamefont
  {Alexeev}, \citenamefont {Candini}, \citenamefont {Coletti}, \citenamefont
  {Turchinovich}, \citenamefont {Kl{\"a}ui}, \citenamefont {Bonn},
  \citenamefont {Koppens}, \citenamefont {Hendry}, \citenamefont {Polini},\
  and\ \citenamefont {Tielrooij}}]{Tomadin:2018}%
  \BibitemOpen
  \bibfield  {author} {\bibinfo {author} {\bibfnamefont {Andrea}\ \bibnamefont
  {Tomadin}}, \bibinfo {author} {\bibfnamefont {Sam~M.}\ \bibnamefont
  {Hornett}}, \bibinfo {author} {\bibfnamefont {Hai~I.}\ \bibnamefont {Wang}},
  \bibinfo {author} {\bibfnamefont {Evgeny~M.}\ \bibnamefont {Alexeev}},
  \bibinfo {author} {\bibfnamefont {Andrea}\ \bibnamefont {Candini}}, \bibinfo
  {author} {\bibfnamefont {Camilla}\ \bibnamefont {Coletti}}, \bibinfo {author}
  {\bibfnamefont {Dmitry}\ \bibnamefont {Turchinovich}}, \bibinfo {author}
  {\bibfnamefont {Mathias}\ \bibnamefont {Kl{\"a}ui}}, \bibinfo {author}
  {\bibfnamefont {Mischa}\ \bibnamefont {Bonn}}, \bibinfo {author}
  {\bibfnamefont {Frank H.~L.}\ \bibnamefont {Koppens}}, \bibinfo {author}
  {\bibfnamefont {Euan}\ \bibnamefont {Hendry}}, \bibinfo {author}
  {\bibfnamefont {Marco}\ \bibnamefont {Polini}}, \ and\ \bibinfo {author}
  {\bibfnamefont {Klaas-Jan}\ \bibnamefont {Tielrooij}},\ }\bibfield  {title}
  {\enquote {\bibinfo {title} {The ultrafast dynamics and conductivity of
  photoexcited graphene at different fermi energies},}\ }\href {\doibase
  10.1126/sciadv.aar5313} {\bibfield  {journal} {\bibinfo  {journal} {Science
  Advances}\ }\textbf {\bibinfo {volume} {4}} (\bibinfo {year} {2018}),\
  10.1126/sciadv.aar5313}\BibitemShut {NoStop}%
\bibitem [{\citenamefont {Al-Naib}\ \emph {et~al.}(2014)\citenamefont
  {Al-Naib}, \citenamefont {Sipe},\ and\ \citenamefont
  {Dignam}}]{Al-Naib:2014}%
  \BibitemOpen
  \bibfield  {author} {\bibinfo {author} {\bibfnamefont {Ibraheem}\
  \bibnamefont {Al-Naib}}, \bibinfo {author} {\bibfnamefont {J.~E.}\
  \bibnamefont {Sipe}}, \ and\ \bibinfo {author} {\bibfnamefont {Marc~M.}\
  \bibnamefont {Dignam}},\ }\bibfield  {title} {\enquote {\bibinfo {title}
  {High harmonic generation in undoped graphene: Interplay of inter- and
  intraband dynamics},}\ }\href {\doibase 10.1103/PhysRevB.90.245423}
  {\bibfield  {journal} {\bibinfo  {journal} {Phys. Rev. B}\ }\textbf {\bibinfo
  {volume} {90}},\ \bibinfo {pages} {245423} (\bibinfo {year}
  {2014})}\BibitemShut {NoStop}%
\bibitem [{\citenamefont {Al-Naib}\ \emph {et~al.}(2015)\citenamefont
  {Al-Naib}, \citenamefont {Poschmann},\ and\ \citenamefont
  {Dignam}}]{Al-Naib:2015}%
  \BibitemOpen
  \bibfield  {author} {\bibinfo {author} {\bibfnamefont {Ibraheem}\
  \bibnamefont {Al-Naib}}, \bibinfo {author} {\bibfnamefont {Max}\ \bibnamefont
  {Poschmann}}, \ and\ \bibinfo {author} {\bibfnamefont {Marc~M.}\ \bibnamefont
  {Dignam}},\ }\bibfield  {title} {\enquote {\bibinfo {title} {Optimizing
  third-harmonic generation at terahertz frequencies in graphene},}\ }\href
  {\doibase 10.1103/PhysRevB.91.205407} {\bibfield  {journal} {\bibinfo
  {journal} {Phys. Rev. B}\ }\textbf {\bibinfo {volume} {91}},\ \bibinfo
  {pages} {205407} (\bibinfo {year} {2015})}\BibitemShut {NoStop}%
\bibitem [{\citenamefont {Winnerl}\ \emph {et~al.}(2011)\citenamefont
  {Winnerl}, \citenamefont {Orlita}, \citenamefont {Plochocka}, \citenamefont
  {Kossacki}, \citenamefont {Potemski}, \citenamefont {Winzer}, \citenamefont
  {Malic}, \citenamefont {Knorr}, \citenamefont {Sprinkle}, \citenamefont
  {Berger}, \citenamefont {de~Heer}, \citenamefont {Schneider},\ and\
  \citenamefont {Helm}}]{Winnerl:2011}%
  \BibitemOpen
  \bibfield  {author} {\bibinfo {author} {\bibfnamefont {S.}~\bibnamefont
  {Winnerl}}, \bibinfo {author} {\bibfnamefont {M.}~\bibnamefont {Orlita}},
  \bibinfo {author} {\bibfnamefont {P.}~\bibnamefont {Plochocka}}, \bibinfo
  {author} {\bibfnamefont {P.}~\bibnamefont {Kossacki}}, \bibinfo {author}
  {\bibfnamefont {M.}~\bibnamefont {Potemski}}, \bibinfo {author}
  {\bibfnamefont {T.}~\bibnamefont {Winzer}}, \bibinfo {author} {\bibfnamefont
  {E.}~\bibnamefont {Malic}}, \bibinfo {author} {\bibfnamefont
  {A.}~\bibnamefont {Knorr}}, \bibinfo {author} {\bibfnamefont
  {M.}~\bibnamefont {Sprinkle}}, \bibinfo {author} {\bibfnamefont
  {C.}~\bibnamefont {Berger}}, \bibinfo {author} {\bibfnamefont {W.~A.}\
  \bibnamefont {de~Heer}}, \bibinfo {author} {\bibfnamefont {H.}~\bibnamefont
  {Schneider}}, \ and\ \bibinfo {author} {\bibfnamefont {M.}~\bibnamefont
  {Helm}},\ }\bibfield  {title} {\enquote {\bibinfo {title} {Carrier relaxation
  in epitaxial graphene photoexcited near the dirac point},}\ }\href {\doibase
  10.1103/PhysRevLett.107.237401} {\bibfield  {journal} {\bibinfo  {journal}
  {Phys. Rev. Lett.}\ }\textbf {\bibinfo {volume} {107}},\ \bibinfo {pages}
  {237401} (\bibinfo {year} {2011})}\BibitemShut {NoStop}%
\bibitem [{\citenamefont {Malic}\ \emph {et~al.}(2011)\citenamefont {Malic},
  \citenamefont {Winzer}, \citenamefont {Bobkin},\ and\ \citenamefont
  {Knorr}}]{Malic:2011}%
  \BibitemOpen
  \bibfield  {author} {\bibinfo {author} {\bibfnamefont {Ermin}\ \bibnamefont
  {Malic}}, \bibinfo {author} {\bibfnamefont {Torben}\ \bibnamefont {Winzer}},
  \bibinfo {author} {\bibfnamefont {Evgeny}\ \bibnamefont {Bobkin}}, \ and\
  \bibinfo {author} {\bibfnamefont {Andreas}\ \bibnamefont {Knorr}},\
  }\bibfield  {title} {\enquote {\bibinfo {title} {Microscopic theory of
  absorption and ultrafast many-particle kinetics in graphene},}\ }\href
  {\doibase 10.1103/PhysRevB.84.205406} {\bibfield  {journal} {\bibinfo
  {journal} {Phys. Rev. B}\ }\textbf {\bibinfo {volume} {84}},\ \bibinfo
  {pages} {205406} (\bibinfo {year} {2011})}\BibitemShut {NoStop}%
\bibitem [{\citenamefont {Winnerl}\ \emph {et~al.}()\citenamefont {Winnerl},
  \citenamefont {Mittendorff}, \citenamefont {König‐Otto}, \citenamefont
  {Schneider}, \citenamefont {Helm}, \citenamefont {Winzer}, \citenamefont
  {Knorr},\ and\ \citenamefont {Malic}}]{Winnerl:2017}%
  \BibitemOpen
  \bibfield  {author} {\bibinfo {author} {\bibfnamefont {Stephan}\ \bibnamefont
  {Winnerl}}, \bibinfo {author} {\bibfnamefont {Martin}\ \bibnamefont
  {Mittendorff}}, \bibinfo {author} {\bibfnamefont {Jacob~C.}\ \bibnamefont
  {König‐Otto}}, \bibinfo {author} {\bibfnamefont {Harald}\ \bibnamefont
  {Schneider}}, \bibinfo {author} {\bibfnamefont {Manfred}\ \bibnamefont
  {Helm}}, \bibinfo {author} {\bibfnamefont {Torben}\ \bibnamefont {Winzer}},
  \bibinfo {author} {\bibfnamefont {Andreas}\ \bibnamefont {Knorr}}, \ and\
  \bibinfo {author} {\bibfnamefont {Ermin}\ \bibnamefont {Malic}},\ }\bibfield
  {title} {\enquote {\bibinfo {title} {Ultrafast processes in graphene: From
  fundamental manybody interactions to device applications},}\ }\href {\doibase
  10.1002/andp.201700022} {\bibfield  {journal} {\bibinfo  {journal} {Ann.
  Phys. (Berlin)}\ }\textbf {\bibinfo {volume} {529}},\ \bibinfo {pages}
  {1700022}}\BibitemShut {NoStop}%
\bibitem [{\citenamefont {Malic}\ \emph {et~al.}(2017)\citenamefont {Malic},
  \citenamefont {Winzer}, \citenamefont {Wendler}, \citenamefont {Brem},
  \citenamefont {Jago}, \citenamefont {Knorr}, \citenamefont {Mittendorff},
  \citenamefont {K\"{o}nig-Otto}, \citenamefont {Plötzing}, \citenamefont
  {Neumaier}, \citenamefont {Schneider}, \citenamefont {Helm},\ and\
  \citenamefont {Winnerl}}]{Malic:2017}%
  \BibitemOpen
  \bibfield  {author} {\bibinfo {author} {\bibfnamefont {E.}~\bibnamefont
  {Malic}}, \bibinfo {author} {\bibfnamefont {T.}~\bibnamefont {Winzer}},
  \bibinfo {author} {\bibfnamefont {F.}~\bibnamefont {Wendler}}, \bibinfo
  {author} {\bibfnamefont {S.}~\bibnamefont {Brem}}, \bibinfo {author}
  {\bibfnamefont {R.}~\bibnamefont {Jago}}, \bibinfo {author} {\bibfnamefont
  {A.}~\bibnamefont {Knorr}}, \bibinfo {author} {\bibfnamefont
  {M.}~\bibnamefont {Mittendorff}}, \bibinfo {author} {\bibfnamefont {J.~C.}\
  \bibnamefont {K\"{o}nig-Otto}}, \bibinfo {author} {\bibfnamefont
  {T.}~\bibnamefont {Plötzing}}, \bibinfo {author} {\bibfnamefont
  {D.}~\bibnamefont {Neumaier}}, \bibinfo {author} {\bibfnamefont
  {H.}~\bibnamefont {Schneider}}, \bibinfo {author} {\bibfnamefont
  {M.}~\bibnamefont {Helm}}, \ and\ \bibinfo {author} {\bibfnamefont
  {S.}~\bibnamefont {Winnerl}},\ }\bibfield  {title} {\enquote {\bibinfo
  {title} {Carrier dynamics in graphene: Ultrafast many‐particle
  phenomena},}\ }\href {\doibase 10.1002/andp.201700038} {\bibfield  {journal}
  {\bibinfo  {journal} {Ann. Phys. (Berlin)}\ }\textbf {\bibinfo {volume}
  {529}},\ \bibinfo {pages} {1700038} (\bibinfo {year} {2017})}\BibitemShut
  {NoStop}%
\bibitem [{\citenamefont {Fang}\ \emph {et~al.}(2011)\citenamefont {Fang},
  \citenamefont {Konar}, \citenamefont {Xing},\ and\ \citenamefont
  {Jena}}]{Fang:2011}%
  \BibitemOpen
  \bibfield  {author} {\bibinfo {author} {\bibfnamefont {Tian}\ \bibnamefont
  {Fang}}, \bibinfo {author} {\bibfnamefont {Aniruddha}\ \bibnamefont {Konar}},
  \bibinfo {author} {\bibfnamefont {Huili}\ \bibnamefont {Xing}}, \ and\
  \bibinfo {author} {\bibfnamefont {Debdeep}\ \bibnamefont {Jena}},\ }\bibfield
   {title} {\enquote {\bibinfo {title} {High-field transport in two-dimensional
  graphene},}\ }\href {\doibase 10.1103/PhysRevB.84.125450} {\bibfield
  {journal} {\bibinfo  {journal} {Phys. Rev. B}\ }\textbf {\bibinfo {volume}
  {84}},\ \bibinfo {pages} {125450} (\bibinfo {year} {2011})}\BibitemShut
  {NoStop}%
\bibitem [{\citenamefont {Razavipour}\ \emph {et~al.}(2015)\citenamefont
  {Razavipour}, \citenamefont {Yang}, \citenamefont {Guermoune}, \citenamefont
  {Hilke}, \citenamefont {Cooke}, \citenamefont {Al-Naib}, \citenamefont
  {Dignam}, \citenamefont {Blanchard}, \citenamefont {Hafez}, \citenamefont
  {Chai}, \citenamefont {Ferachou}, \citenamefont {Ozaki}, \citenamefont
  {L\'evesque},\ and\ \citenamefont {Martel}}]{Razavipour:2015}%
  \BibitemOpen
  \bibfield  {author} {\bibinfo {author} {\bibfnamefont {Hadi}\ \bibnamefont
  {Razavipour}}, \bibinfo {author} {\bibfnamefont {Wayne}\ \bibnamefont
  {Yang}}, \bibinfo {author} {\bibfnamefont {Abdeladim}\ \bibnamefont
  {Guermoune}}, \bibinfo {author} {\bibfnamefont {Michael}\ \bibnamefont
  {Hilke}}, \bibinfo {author} {\bibfnamefont {David~G.}\ \bibnamefont {Cooke}},
  \bibinfo {author} {\bibfnamefont {Ibraheem}\ \bibnamefont {Al-Naib}},
  \bibinfo {author} {\bibfnamefont {Marc~M.}\ \bibnamefont {Dignam}}, \bibinfo
  {author} {\bibfnamefont {Fran\ifmmode \mbox{\c{c}}\else~\c{c}\fi{}ois}\
  \bibnamefont {Blanchard}}, \bibinfo {author} {\bibfnamefont {Hassan~A.}\
  \bibnamefont {Hafez}}, \bibinfo {author} {\bibfnamefont {Xin}\ \bibnamefont
  {Chai}}, \bibinfo {author} {\bibfnamefont {Denis}\ \bibnamefont {Ferachou}},
  \bibinfo {author} {\bibfnamefont {Tsuneyuki}\ \bibnamefont {Ozaki}}, \bibinfo
  {author} {\bibfnamefont {Pierre~L.}\ \bibnamefont {L\'evesque}}, \ and\
  \bibinfo {author} {\bibfnamefont {Richard}\ \bibnamefont {Martel}},\
  }\bibfield  {title} {\enquote {\bibinfo {title} {High-field response of gated
  graphene at terahertz frequencies},}\ }\href {\doibase
  10.1103/PhysRevB.92.245421} {\bibfield  {journal} {\bibinfo  {journal} {Phys.
  Rev. B}\ }\textbf {\bibinfo {volume} {92}},\ \bibinfo {pages} {245421}
  (\bibinfo {year} {2015})}\BibitemShut {NoStop}%
\bibitem [{\citenamefont {Aversa}\ and\ \citenamefont
  {Sipe}(1995)}]{Aversa:1995}%
  \BibitemOpen
  \bibfield  {author} {\bibinfo {author} {\bibfnamefont {Claudio}\ \bibnamefont
  {Aversa}}\ and\ \bibinfo {author} {\bibfnamefont {J.~E.}\ \bibnamefont
  {Sipe}},\ }\bibfield  {title} {\enquote {\bibinfo {title} {Nonlinear optical
  susceptibilities of semiconductors: Results with a length-gauge analysis},}\
  }\href {\doibase 10.1103/PhysRevB.52.14636} {\bibfield  {journal} {\bibinfo
  {journal} {Phys. Rev. B}\ }\textbf {\bibinfo {volume} {52}},\ \bibinfo
  {pages} {14636--14645} (\bibinfo {year} {1995})}\BibitemShut {NoStop}%
\bibitem [{\citenamefont {Ando}(2006)}]{Ando:2006}%
  \BibitemOpen
  \bibfield  {author} {\bibinfo {author} {\bibfnamefont {Tsuneya}\ \bibnamefont
  {Ando}},\ }\bibfield  {title} {\enquote {\bibinfo {title} {Screening effect
  and impurity scattering in monolayer graphene},}\ }\href {\doibase
  10.1143/JPSJ.75.074716} {\bibfield  {journal} {\bibinfo  {journal} {J. Phys.
  Soc. Jpn.}\ }\textbf {\bibinfo {volume} {75}},\ \bibinfo {pages} {074716}
  (\bibinfo {year} {2006})}\BibitemShut {NoStop}%
\bibitem [{\citenamefont {Hwang}\ and\ \citenamefont
  {Das~Sarma}(2008)}]{Hwang:2008}%
  \BibitemOpen
  \bibfield  {author} {\bibinfo {author} {\bibfnamefont {E.~H.}\ \bibnamefont
  {Hwang}}\ and\ \bibinfo {author} {\bibfnamefont {S.}~\bibnamefont
  {Das~Sarma}},\ }\bibfield  {title} {\enquote {\bibinfo {title}
  {Single-particle relaxation time versus transport scattering time in a
  two-dimensional graphene layer},}\ }\href {\doibase
  10.1103/PhysRevB.77.195412} {\bibfield  {journal} {\bibinfo  {journal} {Phys.
  Rev. B}\ }\textbf {\bibinfo {volume} {77}},\ \bibinfo {pages} {195412}
  (\bibinfo {year} {2008})}\BibitemShut {NoStop}%
\bibitem [{\citenamefont {Kuhn}(1998)}]{Kuhn:1998}%
  \BibitemOpen
  \bibfield  {author} {\bibinfo {author} {\bibfnamefont {Tilmann}\ \bibnamefont
  {Kuhn}},\ }\bibfield  {title} {\enquote {\bibinfo {title} {Density matrix
  theory of coherent ultrafast dynamics},}\ }in\ \href@noop {} {\emph {\bibinfo
  {booktitle} {Theory of Transport Properties of Semiconductor
  Nanostructures}}},\ \bibinfo {editor} {edited by\ \bibinfo {editor}
  {\bibfnamefont {Eckehard}\ \bibnamefont {Sch\"{o}ll}}}\ (\bibinfo
  {publisher} {Springer},\ \bibinfo {address} {London},\ \bibinfo {year}
  {1998})\ Chap.~\bibinfo {chapter} {6}, pp.\ \bibinfo {pages}
  {173--214}\BibitemShut {NoStop}%
\bibitem [{\citenamefont {Rossi}\ and\ \citenamefont
  {Kuhn}(2002)}]{Rossi:2002}%
  \BibitemOpen
  \bibfield  {author} {\bibinfo {author} {\bibfnamefont {Fausto}\ \bibnamefont
  {Rossi}}\ and\ \bibinfo {author} {\bibfnamefont {Tilmann}\ \bibnamefont
  {Kuhn}},\ }\bibfield  {title} {\enquote {\bibinfo {title} {Theory of
  ultrafast phenomena in photoexcited semiconductors},}\ }\href {\doibase
  10.1103/RevModPhys.74.895} {\bibfield  {journal} {\bibinfo  {journal} {Rev.
  Mod. Phys.}\ }\textbf {\bibinfo {volume} {74}},\ \bibinfo {pages} {895--950}
  (\bibinfo {year} {2002})}\BibitemShut {NoStop}%
\bibitem [{Gri()}]{Grid}%
  \BibitemOpen
  \href@noop {} {}\bibinfo {note} {We notice a very slight increase in carrier
  density over time due to differences in scattering-in and -out rates as
  calculated via billinear interpolation, and therefore set the gridpoint
  density such that the density changes less than 2\% over the druation of the
  simulation for all field strengths and scattering scenarios.}\BibitemShut
  {Stop}%
\bibitem [{\citenamefont {McGouran}(2016)}]{McGouran:2016}%
  \BibitemOpen
  \bibfield  {author} {\bibinfo {author} {\bibfnamefont {Riley}\ \bibnamefont
  {McGouran}},\ }\emph {\bibinfo {title} {Nonlinear Response of Unbiased and
  Biased Bilayer Graphene at Terahertz Frequencies}},\ \href@noop {} {Master's
  thesis},\ \bibinfo  {school} {Queen's University}, \bibinfo {address}
  {QSpace} (\bibinfo {year} {2016})\BibitemShut {NoStop}%
\end{thebibliography}%

\end{document}